\documentclass[UTF8,12pt,a4paper]{article}
\usepackage{geometry}
\usepackage{graphicx}
\usepackage{epstopdf}
\usepackage{epsfig}
\usepackage{amsmath}
\usepackage{color}
\usepackage{latexsym} 
\usepackage{amssymb}  
\usepackage{slashed}
\usepackage{bm}
\usepackage{hyperref}
\usepackage{CJKutf8}
\usepackage{authblk}
\usepackage{multirow}
\usepackage{booktabs}
\usepackage{verbatim}
\usepackage{epstopdf}
\usepackage{caption,subcaption}
\usepackage[symbol]{footmisc}
\usepackage{lineno}
\begin{document}

\title{A compact analytical approximation for a light sterile neutrino oscillation in matter}
\author[a]{Bao-Biao Yue} 
\author[b,c]{Wei Li}
\author[a]{Jia-Jie Ling \thanks{Corresponding authors: lingjj5@mail.sysu.edu.cn, fanrongxu@jnu.edu.cn}}
\author[b]{Fan-Rong Xu$^{*}$}
\affil[a]{{\tiny School of Physics, Sun Yat-sen University, No. 135, Xingang Xi Road, Guangzhou, 510275, P. R. China}}
\affil[b]{{\tiny Department of physics, Jinan University, No. 601, Huang Pu Road, Guangzhou, 510632, P. R. China}}
\affil[c]{{\tiny Department of Physics and Astronomy, National Taiwan University, No. 1, Sec. 4, Roosevelt Road, Taipei, 10617, P. R. China}}

% other emails
%\ead{yuebb@mail2.sysu.edu.cn}
%\ead{r07222072@ntu.edu.tw}
%\ead{lingjj5@mail.sysu.edu.cn}
\maketitle

\begin{abstract}
The existence of light sterile neutrinos is a long standing question for particle physics. Several experimental ``anomalies'' could be explained by introducing ~eV mass scaled light sterile neutrinos. Many experiments are actively hunting for such light sterile neutrinos through neutrino oscillation. For long baseline experiments, matter effect needs to be treated carefully for precise neutrino oscillation probability calculation.
However, it is usually time-consuming or analytical complexity.  
In this manuscript we adopt the Jacobi-like method to diagonalize the Hermitian Hamiltonian matrix and derive analytically simplified neutrino oscillation probabilities for 3 (active) + 1 (sterile)-neutrino mixing for a constant matter density. These approximations can reach quite high numerical accuracy while keeping its analytical simplicity and fast computing speed. It would be useful for the current and future long baseline neutrino oscillation experiments.

\end{abstract}
%\flushbottom
%\linenumbers
\section{Introduction}
\label{sec:intro}

Neutrino oscillation has been indisputably established by atmospheric, solar, reactor and accelerator experimental results~\cite{cite:PDG}. After the recent reactor experiments~\cite{cite:dayabay,cite:RENO, cite:CHOOZ} discovered the last unknown mixing angle $\theta_{13}$ in the 3-neutrino mixing framework, neutrino oscillation measurement enters a precision era. Nonzero neutrino mass provides convincing evidence of new physics beyond the standard model. Introducing right-handed neutrinos is a natural way to introduce neutrino mass. In the standard electro-weak V-A theory, right-handed neutrinos cannot couple with W$^\pm$ and Z$^0$ bosons. Electron collider experimental data~\cite{cite:lep} constrain the number of active light neutrino flavors to three, other new types of light neutrinos must be sterile. Currently there is no theoretical constraint on sterile neutrino mass. They could be very massive (10$^{15}$~GeV) as suggested by the see-saw mechanism; They also could be dark matter in the keV mass range; And they also might be as light as sub-eV which would explain the CMB measurement. 

If light sterile neutrinos mix with the active neutrinos, their signature could be observed by neutrino oscillation experiments. LSND observed $87.9 \pm 22.4 \pm 6.0$ $\overline{\nu}_e$ signal events from $\overline{\nu}_{\mu}$ source from $\mu^{+}$ decay at rest, which suggests a sterile neutrino with mass greater than 0.4~eV~\cite{cite:lsnd,cite:lsnd1}. Recently MiniBooNE experiments reported a $4.7 \sigma$ excess of electron-like events when combining both the $\nu_\mu$ and $\overline{\nu}_\mu$ beam configurations. The significance of the combined LSND and MiniBooNE excesses can even reach 6$\sigma$~\cite{cite:MiniBooNe2018}, although the source of the low energy excess from MiniBooNE is still unclear. Experimental hints of the existence of eV mass scaled sterile neutrinos also come from short baseline reactor neutrino experiments~\cite{cite:anomaly4, cite:anomaly6, cite:anomaly7}. However, the uncertainties with theoretical reactor antineutrino flux calculation might be underestimated, giving an observed excess of antineutrino events at 4-6~MeV relative to predictions~\cite{cite:RENO_bump,cite:DoubleChooz_bump,cite:DayaBay_bump,cite:DayaBay_bump1,cite:DayaBay_bump2}. Therefore, whether the reactor anomaly is completely cased by the theoretical modeling or sterile neutrinos is still up in the air. 

It is worth mentioning, although eV-scale sterile neutrinos could help to explain several experimental anomalies, they are not quite theoretical motivated. And they are also in tension with the muon neutrino disappearance results, especially for recent results from IceCube~\cite{cite:icecube} and MINOS/MINOS+~\cite{cite:minos+}. The most recent combined analysis with MINOS+, Bugey and Daya Bay experiments set a very strong limit on sterile neutrino mixing~\cite{cite:combinedanalysis}, which can almost completely exclude LSND and MiniBooNE sterile neutrino hypothesis at eV-scale region. However, a more convincing and direct testing would come from muon decay at rest experiments, such as the proposal of JSNS2~\cite{cite:JSNS2}. Certainly, the existence of eV mass scale sterile neutrinos therefore needs further evidence. Many reactor and accelerator neutrino experiments are actively searching for sterile neutrinos at various mass scales~\cite{cite:sterile1,cite:sterile2,cite:sterile3,cite:sterile4,cite:sterile5}. 

For long baseline accelerator neutrino experiments~\cite{cite:NOVA, cite:DUNE}, neutrino matter effect plays an important role in neutrino mass hierarchy~\cite{cite:MH1,cite:MH2} and CP violation~\cite{cite:cp} measurements. As first pointed out by Wolfenstein, neutrinos propagating in matter will oscillate differently from those in a vacuum~\cite{cite:msw1}. The presence of electrons in matter changes the energy levels of propagation eigenstates of neutrinos due to charged current coherent forward scattering of the electron neutrinos. Later on, Mikheyev and Smirnov~\cite{cite:msw2} further noticed the matter effect can produce resonant maximal flavor transition when neutrinos propagate through matter at certain electron densities. Super-Kamiokande observes an indication of different solar neutrino flux during the night and day for solar neutrinos passing through additional terrestrial matter in the earth at different periods~\cite{cite:superK_solar}. For sterile neutrino and other new physics searches, matter effect has to be calculated carefully and precisely, especially for long baseline neutrino oscillation experiments.     

Neutrino oscillation in matter can be solved accurately using numerical or analytical calculation~\cite{cite:previouswork} with a complex matrix diagonalization algorithm. In practice, analytic approximations are more commonly used in neutrino experiments and useful to understand the oscillation features. High precision analytical expressions for 3-neutrino oscillation in matter has been thoroughly studied~\cite{cite:msw_method1,cite:msw_method2,cite:msw_method3,cite:msw_method4,cite:msw_method5,cite:msw_method6,cite:msw_method7,cite:msw_method8,cite:msw_method9,cite:msw_method10,cite:msw_method11,cite:msw_method12,cite:msw_method13,cite:msw_method14}. Some of them utilize perturbation theory and rely on expansions in parameter $\theta_{13}$. Given the large $\theta_{13}$ observed, higher order corrections associated with $\theta_{13}$ are needed to achieve numerical accuracy. Thus the oscillation expressions usually become quite complicated. Ref.~\cite{cite:originalwork} introduces the Jacobi method to diagonalize the real Hermitian matrix. It maintains the same analytic expressions for neutrinos propagating in matter as they have in vacuum in terms of the effective neutrino mixing angles and mass-squared differences in matter. 

For sterile neutrinos, the oscillation expressions will be very complicated if additional light sterile neutrinos exist~\cite{cite:sterilemattereffect}. Compared with standard 3-neutrino mixing, the simplest 3 (active) + 1 (sterile)-neutrino mixing has 3 additional mixing angles (i.e. $\theta_{14}$, $\theta_{24}$ and $\theta_{34}$) and 2 additional CP phases (i.e. $\delta_{24}$ and $\delta_{34}$). Furthermore, since sterile neutrinos do not interact with matter, the neutral current potential for active neutrinos also needs to be taken into account.
N. Klop et. al.~\cite{cite:Klop} provides a method to covert 3+1-neutrino mixing with matter effects into a Non-Standard Interaction (NSI) problem in the 3-neutrino mixing case. 
Here we follow the rotation strategy introduced in Ref.~\cite{cite:originalwork} and adopt the Jacobi-like method~\cite{cite:jacobi-like1,cite:jacobi-like2}, which is able to diagonalize the Hermitian complex matrix, to derive analytical approximations for the 3+1-neutrino oscillation in matter. While keeping the formula simplicity, the expressions can also achieve very good numerical accuracy and fast calculation speed. This could be very useful for the current and near future neutrino oscillation experiments.

This paper starts with the section~\ref{sec:NeuTheo} and introduces the fundamental theory of neutrino mixing and oscillation, including sterile neutrinos and matter effect. The basic idea of the Jacobi-like method and the derivation of analytical approximations for sterile neutrino oscillation probabilities are presented in section~\ref{sec:Appro}. In the end, the accuracy of this work is shown in section~
\ref{sec:accuracy} with two long baseline accelerator neutrino experiments as demonstrations. More details about the Jacobi-like method and formula derivation are listed in the appendix.

\section{Theoretical framework}
\label{sec:NeuTheo}
\subsection{Neutrino oscillation}
\label{sec:NeuTheo:Osci}
In the standard neutrino mixing paradigm, three neutrino flavor eigenstates ($\nu_e$, $\nu_\mu$, $\nu_\tau$) are
superpositions of three neutrino mass eigenstates ($\nu_1$, $\nu_2$, $\nu_3$).
\begin{equation}
\label{eq:superposition}
 \begin{pmatrix}
 \nu_e\\
 \nu_\mu \\
 \nu_\tau
 \end{pmatrix}
 =U
 \begin{pmatrix}
  \nu_1 \\
  \nu_2 \\
  \nu_3
 \end{pmatrix}\,.
\end{equation}
Here $U$ is the so-called PMNS (Pontecorvo-Maki-Nakawaga-Sakata) mixing matrix \cite{cite:PMNS1,cite:PMNS2,cite:PMNS3},
which can be parametrized as
\begin{equation}
\label{eq:3mixing}
\begin{split}
U=&R_{23}(\theta_{23},0)R_{13}(\theta_{13},\delta_{13})R_{12}(\theta_{12},0)
\\
=&
\begin{bmatrix}
 1&0&0\\
 0&c_{23}&s_{23}\\
 0&-s_{23}&c_{23}
\end{bmatrix}
\begin{bmatrix}
 c_{13}&0&s_{13}e^{-i\delta_{13}}\\
 0&1&0\\
 -s_{13}e^{i\delta_{13}}&0&c_{13}
\end{bmatrix}
\begin{bmatrix}
 c_{12}&s_{12}&0\\
 -s_{12}&c_{12}&0\\
 0&0&1
\end{bmatrix}
\end{split}  \,,
\end{equation}
\noindent where $R_{ij}(\theta_{ij},\delta_{ij})$ denotes a counterclockwise rotation in the complex $ij$-plane through a mixing angle $\theta_{ij}$ and a CP phase $\delta_{ij}$ with $c_{ij}=\cos\theta_{ij}$ and $s_{ij}=\sin\theta_{ij}$.
This work adopts the conventions $0\le\theta_{ij}\le\pi/2$ and $0\le\delta_{ij}\le 2\pi$.

Under the plane wave assumption, the general oscillation probability from $\alpha$-flavor type neutrinos to $\beta$-flavor type neutrinos can be expressed as
\begin{equation}
\label{eq:ProVacu}
\begin{split}
 P_{\nu_\alpha \rightarrow \nu_\beta}
=\delta_{\alpha\beta} & -4\sum\limits_{i>j}
 \Re \left (U_{\beta i} U_{\alpha i}^* U_{\beta j}^* U_{\alpha j}\right)\sin ^2\Delta_{ij}
\\&
\pm 2\sum\limits_{i>j}\Im \left (U_{\beta i} U_{\alpha i}^* U_{\beta j}^* U_{\alpha j}\right)\sin 2\Delta_{ij}
 \end{split} \,,\qquad(i,j=1,2,3)
\end{equation}
\noindent where the upper and lower sign is for the neutrino and antineutrino cases respectively. 
$\Delta_{ij}$ stands for
\begin{equation}
\label{eq:Delta}
 \Delta_{ij}\equiv\frac{\Delta m^2_{ij}L}{4E}
 =1.267\left(\frac{\Delta m^2_{ij}}{{\rm eV}^2}\right)\left(\frac{\rm GeV}{E}\right)\left(\frac{L}{\rm km}\right)\,,
\end{equation}
\noindent where $\Delta m^2_{ij} = m^2_i-m^2_j$ is the mass-squared difference between neutrino mass eigenstates $\nu_{i}$ and $\nu_{j}$. 

According to eq.~\eqref{eq:3mixing} and eq.~\eqref{eq:ProVacu}, 3-flavor neutrino oscillation is described with six parameters, including two independent neutrino mass squared differences ($\Delta m_{21}^2$ and $\Delta m_{32}^2$), three mixing angles ($\theta_{12}$, $\theta_{13}$ and $\theta_{23}$) and one leptonic CP phase ($\delta_{13}$). Following the same convention, the 4-flavor neutrino mixing matrix can be parametrized as
\begin{equation}\label{eq:4mixing}
U=R_{34}(\theta_{34},\delta_{34})
R_{24}(\theta_{24},\delta_{24})R_{14}(\theta_{14},0)R_{23}(\theta_{23},0)
R_{13}(\theta_{13},\delta_{13})R_{12}
(\theta_{12},0) \,,
\end{equation}
\noindent with six additional neutrino oscillation parameters:
$\theta_{14}$, $\theta_{24}$, $\theta_{34}$, $\delta_{24}$, $\delta_{34}$ and $\Delta m_{41}^2$ 
\footnote{This is equivalent to use $\delta_{14}$ and $\delta_{24}$, or $\delta_{14}$ and $\delta_{34}$ for the additional CP phases.}.
The exact parameterization expression for each mixing element is listed in the appendix~\ref{sec:appendixA}. 
The general expression for the neutrino oscillation probabilities still follow eq.~\eqref{eq:ProVacu} by simply increasing the total number of neutrino flavors and mass eigenstates to 4. 

In practice, when sterile neutrinos are much heavier than active neutrinos ($|\Delta m_{41}^2| \gg |\Delta m_{31}^2|$), due to finite detector space and energy resolution, 
the rapid oscillation frequency associated with large mass-squared differences between the 4th and the other mass eigenstates 
$\Delta m_{4k}^2 (k=1,2,3)$ will be averaged out, leading to
$\langle\sin^2\Delta_{4k} \rangle \approx \frac{1}{2}$.
The neutrino oscillation equation can then be simplified to
\begin{equation}
\begin{split}
\label{eq:EffPro1}
P_{\nu_\alpha \rightarrow \nu_\beta }
=&
\delta_{\alpha\beta}
-4\sum\limits_{i>j}\Re \left (U_{\beta i}
U_{\alpha i}^* U_{\beta j}^*
U_{\alpha j}\right)\sin ^2\Delta_{ij}
\\&
\pm2\sum\limits_{i>j}\Im \left (U_{\beta i}
U_{\alpha i}^* U_{\beta j}^*
U_{\alpha j}\right)\sin 2\Delta_{ij}
\\
& - \frac{1}{2}\sin^22\theta_{\alpha \beta}
\end{split}
 \qquad(i,j=1,2,3)
\end{equation}
\noindent with $\sin^2 2\theta_{\alpha \beta}=4|U_{\alpha4}|^2(\delta_{\alpha\beta}-|U_{\beta4}|^2)$.
In this paper we prefer to use the full oscillation formula to preserve the rapid oscillation induced by sterile neutrinos.

%Moreover, the neutrino disappearance oscillation probability can be written as

%\begin{equation}
%\begin{split}
%\label{eq:EffPro2}
%P_{\nu_\alpha \rightarrow \nu_\alpha }
%=1 &
%-4\sum\limits_{i>j}|U_{\alpha i}|^2
% |U_{\alpha j}|^2
% \sin ^2\Delta_{ij}
% -\frac{1}{2}\sin^22\theta_{\alpha \alpha}
%\end{split}\,,
%\qquad(i,j=1,2,3)
%\end{equation}

%\noindent with $\sin^22\theta_{\alpha \alpha} =4|U_{\alpha4}|^2(1-|U_{\alpha4}|^2)\approx 4|U_{\alpha4}|^2$ in the case of $|U_{\alpha 4}|^2 \ll 1$.

\subsection{Matter effect}
\label{sec:NeuTheo:MatterEffect}
When active neutrinos propagate through matter, the evolution equation is modified by coherent interaction potentials,
which are generated through coherent forward elastic weak charged-current (CC) and the neutral-current (NC) scattering in a medium.
All active neutrinos can interact with electrons, neutrons and protons in matter through the exchange of a $Z$ boson in the NC process.
However, only electron neutrinos participate in the CC process with electrons through the exchange of $W^{\pm}$.

For electron neutrinos, CC potential is proportional to electron number density. $V_{\mathrm{CC}}=\sqrt{2}G_{F}N_{e}$, where $G_{F}$ is the Fermi coupling constant, $N_{e}$ is the electron number density. The NC potentials caused by electrons and protons will cancel each other because they have opposite signs and the number densities of electrons and protons are basically the same in the earth. The net NC potential, $V_{\mathrm{NC}}=-\frac{\sqrt{2}}{2}G_{F}N_{n}$, is only sensitive to the neutron number density, $N_{n}$. Both $V_{\mathrm{CC}}$ and $V_{\mathrm{NC}}$ need to swap signs for antineutrinos.

For 3-flavor neutrino oscillation, only CC potential needs to be considered for the electron neutrino eigenstate, while the NC potential is a common term for all neutrino flavors and has no net effect on neutrino oscillation.
However, the NC potential cannot be neglected in 3+1-flavor neutrino case, 
since sterile neutrinos do not interact with matter.
The effective Hamiltonian in the flavor eigenstate representation for 3+1-flavor neutrino mixing is
\begin{equation}
\label{eq:NeuH}
\mathcal{H}=\mathcal{H}_v + V =
\frac{1}{2E}
\left(
U\begin{bmatrix}
0&0&0&0\\
0&\Delta m_{21}^2&0&0\\
0&0&\Delta m_{31}^2&0\\
0&0&0&\Delta m_{41}^2
\end{bmatrix}U^{\dag}
+
\begin{bmatrix}
A_{\mathrm{CC}}&0&0&0\\
0&0&0&0\\
0&0&0&0\\
0&0&0&A_{\mathrm{NC}}
\end{bmatrix}
\right)
 \,,
\end{equation}
\noindent where $\mathcal{H}_{v}$ is the
neutrino Hamiltonian in vacuum and $V$ is the matter effect potential. $A_{\mathrm{CC}}$ and $A_{\mathrm{NC}}$ for neutrinos are given by
\begin{subequations}\label{eq:A}
\begin{gather}
\label{eq:Acc}
A_{\mathrm{CC}}=2EV_{\mathrm{CC}}=7.63\times10^{-5}({\rm eV}^2)(\frac{\rho}
{{\rm g/cm}^3})(\frac{E}{\rm GeV}) \,,
\\
\label{eq:Anc}
A_{\mathrm{NC}}=-2EV_{\mathrm{NC}}=3.815\times10^{-5}({\rm eV}^2)(\frac{\rho}
{{\rm g/cm}^3})(\frac{E}{\rm GeV}) \,,
\end{gather}
\end{subequations}
respectively, where $\rho$ is the mass density. 
Similarly to $V_{\mathrm{CC}}$ and $V_{\mathrm{NC}}$, both $A_{\mathrm{CC}}$ and $A_{\mathrm{NC}}$ have to swap signs for antineutrinos. 
%According to the second term in eq.~\eqref{eq:NeuH}, only two rows ($U_{ei}$, $U_{si}$ $(i=1,2,3,4)$)
%elements in $U$ are involved along with $A_{CC}$ and $A_{NC}$ in the Hamiltonian.\\
In this work, we assume a constant $\rho$. If there is no special declaration, $\rho$ will be set to $2.6$ g/cm$^3$ as default.

The evolution of neutrino flavor state $\Psi_{\alpha}$ can be calculated using Schr$\ddot{\rm o}$dinger equation 
$i\frac{d}{dt}\Psi_{\alpha} =  \mathcal{H}\Psi_{\alpha}$.
After diagonalizing the effective Hamiltonian matrix $\mathcal{H}$, we can calculate neutrino oscillation probability in matter 
through the equation
\begin{equation}
\begin{split}
\label{eq:EffPro}
P_{\nu_\alpha \rightarrow \nu_\beta }
=\delta_{\alpha\beta}&
-4\sum\limits_{i>j}\Re \left (\widetilde{U}_{\beta i}
\widetilde{U}_{\alpha i}^* \widetilde{U}_{\beta j}^*
\widetilde{U}_{\alpha j}\right)\sin ^2\widetilde{\Delta}_{ij}
\\&
\pm2\sum\limits_{i>j}\Im \left (\widetilde{U}_{\beta i}
\widetilde{U}_{\alpha i}^* \widetilde{U}_{\beta j}^*
\widetilde{U}_{\alpha j}\right)\sin 2\widetilde{\Delta}_{ij}
\end{split}\,,
 \qquad(i,j=1,2,3,4)
\end{equation}
with the effective mixing matrix $\widetilde{U}$ and effective mass-squared differences $\Delta\widetilde{m}_{ij}^2(i,j=1,2,3,4)$.
In the following approximations, we will rotate the Hamiltonian from mass eigenstate. For simplicity, write the effective Hamiltonian in mass eigenstate as
\begin{equation}
 \label{eq:H}
 H=U^\dag \mathcal{H}U=
 \frac{1}{2E}
\begin{bmatrix}
H_{11}&H_{12}&H_{13}&H_{14} \\
H_{21}&H_{22}&H_{23}&H_{24}\\
H_{31}&H_{32}&H_{33}&H_{34}\\
H_{41}&H_{42}&H_{43}&H_{44}\\
\end{bmatrix} \,,
\end{equation}
\noindent where the Hermitian matrix element $H_{ij}$ yields 
\begin{equation}\label{eq:HElements}
H_{ij}=
\begin{cases}
A_{\mathrm{CC}}U_{ei}^*U_{ej} + A_{\mathrm{NC}}U_{si}^*U_{sj} & (i \neq j) \\
\Delta m_{i1}^2 + A_{\mathrm{CC}}|U_{ei}|^2 + A_{\mathrm{NC}}|U_{si}|^2 & (i=j)
\end{cases} \,.
\end{equation}
In this case, the effective mixing $\widetilde{U}$ yields $\widetilde{U}=UR$, in which $R$ is the diagonalization matrix on $H$.
\section{The analytical approximation}
\label{sec:Appro}
As shown in ref.~\cite{cite:previouswork}, the exact solution for the effective mixing matrix $\widetilde{U}$ and effective mass-squared differences $\Delta\widetilde{m}_{ij}^2(i,j=1,2,3,4)$ can be obtained analytically. However, to obtain higher precision analytical approximations for neutrino oscillation in matter would be more convenient and time-saving. Here we would like to introduce a Jacobi-like method, which is a unitary transformation operation method to diagonalize the complex Hermitian matrix. Then we present the effective mixing matrix and effective mass-squared differences of the 3+1-flavor neutrino mixing framework for both neutrinos and antineutrinos. As a result, high accuracy can be obtained for the calculation of neutrino oscillation probabilities in matter.

\subsection{Jacobi-like method: Diagonalization of a 2 \texorpdfstring{$\times$}{} 2 Hermitian matrix}
 \label{sec:Jacobi}
The Jacobi-like method, which originates from the Jacobi eigenvalue algorithm,
is a effective matrix rotation approach to a diagonalize complex Hermitian matrix. Here we start with an example of solving a 2 $\times$ 2 Hermitian matrix.
A Hermitian matrix
\begin{equation}
\label{eq:M}
M=
\begin{bmatrix}
\alpha&\beta\\
\beta^*&\gamma
\end{bmatrix}
 \qquad (\alpha\,,\gamma \in \mathbb{R} \,,\quad \beta \in \mathbb{C})
\end{equation}
 can be diagonalized as 
 \begin{equation}
\label{eq:MPrime}
M^\prime=R^{\dag}(\omega,\phi)MR(\omega,\phi)=
%\begin{bmatrix}
%\cos\omega & -\sin\omega e^{-i\phi}\\
%\sin\omega e^{i\phi} & \cos\omega
%\end{bmatrix}
%\begin{bmatrix}
%\alpha&\beta\\
%\beta^*&\gamma
%\end{bmatrix}
%\begin{bmatrix}
%\cos\omega & \sin\omega e^{-i\phi}\\
%-\sin\omega e^{i\phi} & \cos\omega
%\end{bmatrix}
%\\=
\begin{bmatrix}
\lambda_-&0\\
0&\lambda_+
\end{bmatrix}
\end{equation}
with a rotation matrix
\begin{equation}
\label{eq:R}
R(\omega,\phi)=
\begin{bmatrix}
\cos\omega & \sin\omega e^{-i\phi}\\
-\sin\omega e^{i\phi} & \cos\omega
\end{bmatrix} \,, \qquad (\omega, \phi \in \mathbb{R})
\end{equation}
\noindent where $\phi=\mathrm{Arg}(\mathrm{sign}(A)\beta^*)$, $ A=\pm|\beta|$ and $ \tan \omega=\frac{2A}{\gamma-\alpha \pm \sqrt{(\gamma-\alpha)^2+4A^2}}$.
The choice of $\pm$ sign for $A$ is optional. 
For simplicity, we choose it to be the same sign as $A_{\mathrm{CC}}$ and $A_{\mathrm{NC}}$ in eq.~\eqref{eq:A} for matter effect in the 3+1 framework.
The $\pm$ sign in the denominator of $\tan\omega$ is correlated with
the exchange of the values of $\lambda_-$ and $\lambda_+$ in eq.~\eqref{eq:MPrime}.
%\footnote{Note that we keep the order of $\lambda_-$ and $\lambda_+$ as that of $\alpha$ and $\lambda$ when $A$ tends to be 0.}
In this work we adopt $+$($-$) for the i-j submatrix diagonalization if $\Delta m_{ij}^2>0$ ($\Delta m_{ij}^2<0$).
After rotation, the eigenvalues of $M$ can be obtained as
\begin{equation}\label{eq:Lambda}
\lambda_-=\frac{\alpha+\gamma \tan^2\omega-2A\tan\omega}{1+\tan^2\omega} \,,\quad
\lambda_+=\frac{\alpha \tan^2\omega+\gamma+2A\tan\omega}{1+\tan^2\omega} \,.
\end{equation}
In a summary, this method is easily used to diagonalize a complex Hermitian matrix through rotation, in which the complex factor $\phi$ is used to deal with the complex diagonalization.

\subsection{The application of Jacobi-like method on 3+1-flavor neutrino mixing}
\label{sec:Appro:EffUM}
To accurately diagonalize the $4\times4$ neutrino Hamiltonian Hermitian matrix using the Jacobi-like method, in principle, we need to perform infinite iterations of $2 \times 2$ submatrix rotation. However, in practice, with only two continuous rotations on the effective Hamiltonian, we already can get analytical approximations for neutrino oscillation in matter with very high accuracy. 
The diagonalized Hamiltonian yields 
\begin{equation}
\hat{H}=R^{\dagger}HR\approx R^{2,\dagger} R^{1,\dagger}HR^{1} R^{2}
%%=R^{2,\dagger} R^{1,\dagger} U^\dagger ( U\mathcal{H}_vU^\dagger  + V) U R^{1} R^{2}
=\widetilde{U}^\dagger ( U\mathcal{H}_vU^\dagger  + V)\widetilde{U}\,,
\end{equation}
where $\widetilde{U}=U R^{1} R^{2}$, and $R^1$ and $R^2$ are the rotation matrices. 
After some mathematical simplifications, $\widetilde{U}$ can be expressed as $R_{34}R_{24}R_{14}R_{23}R_{13}R_{12}$, which has the same form as standard neutrino mixing $U$.
For simplicity, we just show the major results of $\widetilde{U}$ and $\Delta \widetilde{m}_{ij}^2(i,j=1,2,3,4)$ in this section. The complete derivations are shown in appendix~\ref{sec:NeutrinoCase} and \ref{sec:AntiNeutrinoCase}.    

With two continuous rotations on the effective Hamiltonian $H$, we can obtain the effective neutrino mixing matrix $\widetilde{U}$
\begin{equation}
\label{eq:EffU}
\widetilde{U}
\approx
R_{34}(\theta_{34},\delta_{34})
R_{24}(\theta_{24},\delta_{24})R_{14}(\theta_{14},0)R_{23}(\theta_{23},0)
R_{13}(\widetilde{\theta}_{13},\widetilde{\delta}_{13})R_{12}
(\widetilde{\theta}_{12},\widetilde{\delta}_{12}).
\end{equation}
It is very similar to the one in vacuum (i.e eq.~\eqref{eq:4mixing}), 
except that there is one additional effective phase $\widetilde{\delta}_{12}$ in the submatrix $R_{12}$. 
$\widetilde{\theta}_{12}$, $\widetilde{\theta}_{13}$, $\widetilde{\delta}_{13}$
and $\widetilde{\delta}_{12}$ are the effective angles and phases as functions of $E$ in $R_{13}$ and $R_{12}$.
And $R_{34}$, $R_{24}$, $R_{14}$ and $R_{23}$ are the same as in vacuum.
In the diagonalization process, it is always better to first apply a rotation to the submatrix which has the largest absolute ratio of the off-diagonal element to the difference of diagonal ones. Since $\Delta m_{21}^2$ is the smallest mass-squared difference compared with others, we can start with $R_{12}$ submatrix rotation first. 

After the first rotation with $R^1=R_{12}({\omega_1},{\phi_1})$ submatrix, we can obtain the effective angle $\widetilde{\theta}_{12}$ and effective phase $\widetilde{\delta}_{12}$ represented as functions of ${\omega_1}$ and ${\phi_1}$ through the combination of $R_{12}(\theta_{12},0)R_{12}({\omega_1},\phi_1)$: 
\begin{subequations}\label{eq:EffU12}
\begin{gather}
\label{eq:EffU12:1}
 \sin\widetilde{\theta}_{12}
 \approx \frac{|c_{12}\tan{\omega_1} e^{i{\phi_1}}+s_{12}|}{\sqrt{1+\tan^2{\omega_1}}}
 \,,\quad
 \cos\widetilde{\theta}_{12}
 \approx \frac{|c_{12}-s_{12}\tan{\omega_1} e^{i{\phi_1}}|}{\sqrt{1+\tan^2{\omega_1}}}
 \,,
 \\
\label{eq:EffU12:3}
e^{i\widetilde{\delta}_{12}}
\approx
\frac{(c_{12}\tan{\omega_1} e^{i{\phi_1}}+s_{12})(c_{12}-s_{12}\tan{\omega_1} e^{-i{\phi_1}})}
{\cos\widetilde{\theta}_{12}\sin\widetilde{\theta}_{12}(1+\tan^2{\omega_1})}
\,,
\end{gather}
\end{subequations}
in which $\tan{\omega_1}=\frac{2A_{{\omega_1}}}{(H_{22}-H_{11})+
\sqrt{(H_{22}-H_{11})^2+4A_{{\omega_1}}^2}}$, $A_{{\omega_1}}=\pm|H_{12}|$ and $\phi_1=\mathrm{Arg}(\mathrm{sign}(A_{\omega_1})H_{12}^{*})$.
The $+$ and $-$ signs in $A_{\omega_1}$ are for the neutrino and antineutrino cases respectively.
After the first rotation (\eqref{eq:NeuHPrime} and \eqref{eq:AntiNeuHPrime}), we can obtain the eigenvalues of the effective Hamiltonian submatrix
\begin{equation}\label{eq:EffU12:lambda}
\lambda_-=\frac{H_{11}+H_{22}\tan^2{\omega_1}-2A_{\omega_1}\tan{\omega_1} }
{1+\tan^2{\omega_1}} \,,\quad
\lambda_+=\frac{H_{11}\tan^2{\omega_1}+H_{22}+2A_{\omega_1}\tan{\omega_1} }
{1+\tan^2{\omega_1}} \,.
\end{equation}

After partial diagonalization on the 1-2 submatrix, the off-diagonal elements of 1-3 and 2-3
submatrices become the relatively largest of the rest of the submatrices for both neutrinos and antineutrinos cases due to the smallness of the sterile neutrinos mixing angles (i.e. $\theta_{14}$, $\theta_{24}$, $\theta_{34}$). 
In the second rotation, we adopt $R^2=R_{23}({\omega_2},{\phi_2})$ ($R^2=R_{13}({\omega_2},{\phi_2})$)
\footnote{Rotation is chosen by considering convenience of calculations shown in \ref{sec:NeutrinoCaseAbsorpation} and \ref{sec:AntiNeutrinoCaseAbsorpation}.} 
rotation matrix for the neutrino (antineutrino) case. 
After the second rotation, we obtain the $\widetilde{\theta}_{13}$ and $\widetilde{\delta}_{13}$ as the functions of $\omega_2$ and $\phi_2$:
\begin{subequations}\label{eq:EffU13}
\begin{gather}
\label{eq:EffU13:1}
\sin\widetilde{\theta}_{13} \approx
\frac{|c_{13}\tan{\omega_2} e^{i{\phi_2}}+s_{13}e^{i\delta_{13}}|}
{\sqrt{1+\tan^2{\omega_2}}}
\,,\quad
\cos\widetilde{\theta}_{13} \approx
\frac{|c_{13}-s_{13}\tan{\omega_2} e^{i(\delta_{13}-{\phi_2})}|}
{\sqrt{1+\tan^2{\omega_2}}}
\,,
\\
\label{eq:EffU13:3}
e^{i\widetilde{\delta}_{13}}
\approx
\frac{(c_{13}\tan{\omega_2} e^{i{\phi_2}}+s_{13}e^{i\delta_{13}})(c_{13}-s_{13}
\tan{\omega_2} e^{i(\delta_{13}-{\phi_2})})}
{\cos\widetilde{\theta}_{13}\sin\widetilde{\theta}_{13}(1+\tan^2{\omega_2})}
\,,
\end{gather}
\end{subequations}
in which $\tan{\omega_2}=
\frac{2A_{\omega_2}}{(H_{33}-\lambda_{\pm})\pm
\sqrt{(H_{33}-\lambda_{\pm})^2+4A_{\omega_2}^2}}$.
In the equation for $\tan{\omega_2}$, the upper (lower) sign in front of $\sqrt{(H_{33}-\lambda_{\pm})^2+4A_{\omega_2}^2}$
is for NH (IH) (i.e. normal hierarchy (inverted hierarchy)) case, and  $\lambda_+$ ($\lambda_-$) is for neutrino (antineutrino) case.
In the above equations, $A_{\omega_2}$ and $e^{i{\phi_2}}$ have different expressions for neutrinos and antineutrinos. 
For the neutrino case, 
\begin{equation}\label{eq:EffU13:amplitude}
A_{{\omega_2}}=|H_{23}^\prime| \,,
\quad 
\phi_2=\mathrm{Arg}(\mathrm{sign}(A_{\omega_2})H_{23}^{\prime*})
\,,
\quad
H_{23}^\prime=
\frac{H_{13}\tan{\omega_1} e^{i\phi_1}+H_{23}}{\sqrt{1+\tan^2{\omega_1}}}
\,.
\end{equation}
While for the antineutrino case,
\begin{equation}\label{eq:EffU:element}
A_{{\omega_2}}=-|H_{13}^\prime| \,,
\quad 
\phi_2=\mathrm{Arg}(\mathrm{sign}(A_{\omega_2})H_{13}^{\prime*})
\,,
\quad
H_{13}^\prime=\frac{H_{13}-H_{23}\tan{\omega_1} e^{-i\phi_1}}{\sqrt{1+\tan^2{\omega_1}}}
\,.
\end{equation}
In this rotation, we can diagonalize 2-3 (1-3) submatrix for neutrinos (antineutrinos) in eq.~\eqref{eq:NeuHPrimePrime} 
(eq.~\eqref{eq:AntiNeuHPrimePrime}), resulting in two eigenvalues $\lambda_{\pm}^\prime$. 
The formula for $\lambda_{\pm}^\prime$ are
\begin{equation}\label{eq:EffU13:lambda}
\lambda_-^\prime=
\frac{\lambda_+ + H_{33}\tan^2{\omega_2}-2A_{\omega_2}\tan{\omega_2} }
{1+\tan^2{\omega_2}}
\,,\quad
\lambda_+^\prime=
\frac{\lambda_+\tan^2{\omega_2} + H_{33}+2A_{\omega_2}\tan{\omega_2} }
{1+\tan^2{\omega_2}}
\,.
\end{equation}
for the neutrino case, and 
\begin{equation}\label{eq:AntiEffU13:lambda}
\lambda_-^\prime=
\frac{\lambda_- + H_{33}\tan^2{\omega_2}-2A_{\omega_2}\tan{\omega_2} }
{1+\tan^2{\omega_2}}
\,,\quad
\lambda_+^\prime=
\frac{\lambda_-\tan^2{\omega_2}+H_{33}+2A_{\omega_2}\tan{\omega_2} }
{1+\tan^2{\omega_2}}
\,.
\end{equation}
for the antineutrino case.

When the mixing between sterile neutrinos and active neutrinos is relatively small
and neutrino beam energy is $E<100$ GeV,
the off-diagonal elements in the effective Hamiltonian will be very small compared with the diagonal ones
after two of the above rotations are performed. Namely the effective Hamiltonian is approximately diagonalized.
%$\widetilde{U}$ in eq.~\eqref{eq:EffU} can be simplified as $UR_{12}R_{23}$ ($UR_{12}R_{13}$) for neutrino (antineutrino) case. 
So far, all of the effective parameters 
(i.e. $\widetilde{\theta}_{12}$, $\widetilde{\delta}_{12}$, $\widetilde{\theta}_{13}$ 
and $\widetilde{\delta}_{13}$) in $\widetilde{U}$ are presented. 
The diagonal terms in the effective Hamiltonian in the new representation can be treated as $\widetilde{m}_i^2\,(i=1,2,3,4)$. 
After subtracting the smallest neutrino (antineutrino) mass $\lambda_-$ ($\lambda_-^\prime$), we can get the effective neutrino (antineutrino) mass-squared difference $\Delta\widetilde{m}_{ij}^2$ as
\begin{equation}\label{eq:EffM}
\Delta \widetilde{m}_{21}^2 \approx \lambda_{-}^{\prime}-\lambda_-
\,,\quad
\Delta \widetilde{m}_{31}^2 \approx \lambda_{+}^{\prime}-\lambda_-
\,,\quad
\Delta \widetilde{m}_{41}^2 \approx H_{44}-\lambda_-
\,.
\end{equation}
for the neutrino case, and 
\begin{equation}\label{eq:AntiEffM}
\Delta \widetilde{m}_{21}^2 \approx \lambda_+-\lambda_-^\prime
\,,\quad
\Delta \widetilde{m}_{31}^2 \approx \lambda_{+}^{\prime}-\lambda_-^\prime
\,,\quad
\Delta \widetilde{m}_{41}^2 \approx H_{44}-\lambda_-^\prime
\,.
\end{equation}
for the antineutrino case.
 
\begin{figure}[htbp]
\centering
\includegraphics[width=\textwidth]{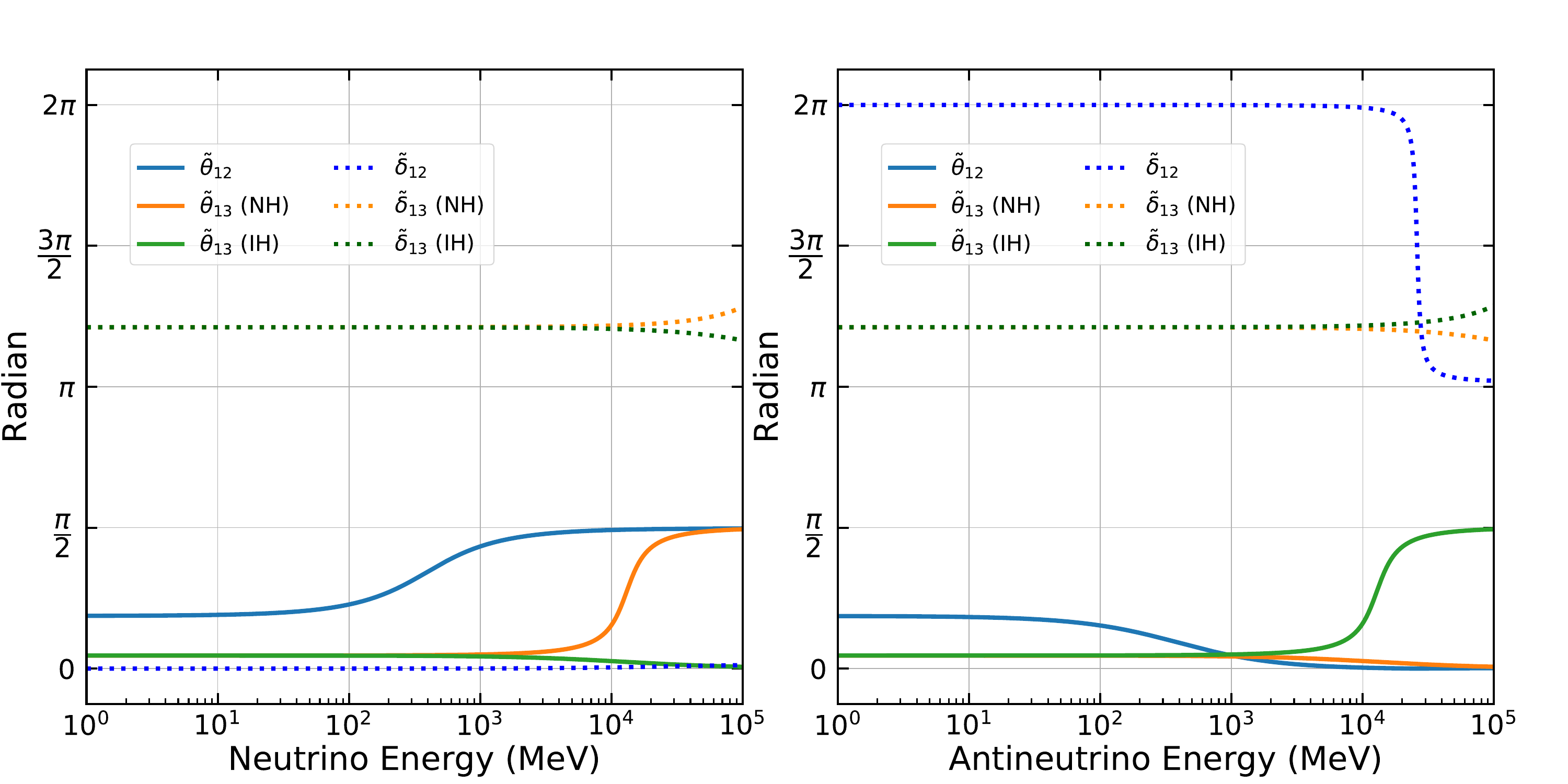}
\caption{\label{fig:1} The values of $\widetilde{\theta}_{12}$, $\widetilde{\theta}_{13}$, $\widetilde{\delta}_{12}$ and $\widetilde{\delta}_{13}$ with respect to neutrino energy. 
In this figure, we assume $\Delta m_{41}^2=0.1$ eV$^2$, $\sin^2\theta_{14}=0.019$, $\sin^2\theta_{24}=0.015$, $\sin^2\theta_{34}=0$~\cite{cite:globalfit},
 $\delta_{13}=218^\circ$~\cite{cite:Neutrino2018} and $\delta_{24}=\delta_{34}=0^\circ$.
The solid and dashed lines are the effective angles and phases respectively. 
The shift of $\widetilde{\delta}_{12}$ with a factor of $\pi$ is caused by the transmission of the sign “-" from $\sin\widetilde{\theta}_{12}$ and $\cos\widetilde{\theta}_{12}$, in which we set $\widetilde{\theta}_{12}$ within $[0,\frac{\pi}{2}]$, in eq.\eqref{eq:EffU12}. As the energy rising in the case of the right plot, 
$\widetilde{\theta}_{12}$ tends to be negative. At that time, we shift the negative sign of it to the phase $\widetilde{\delta}_{12}$ to make sure that $\widetilde{\theta}_{12}$ is in $[0,\frac{\pi}{2}]$, resulting in a $\pi$ shift.
 } 
\end{figure}

\begin{figure}[htbp]
\centering
\includegraphics[width=\textwidth]{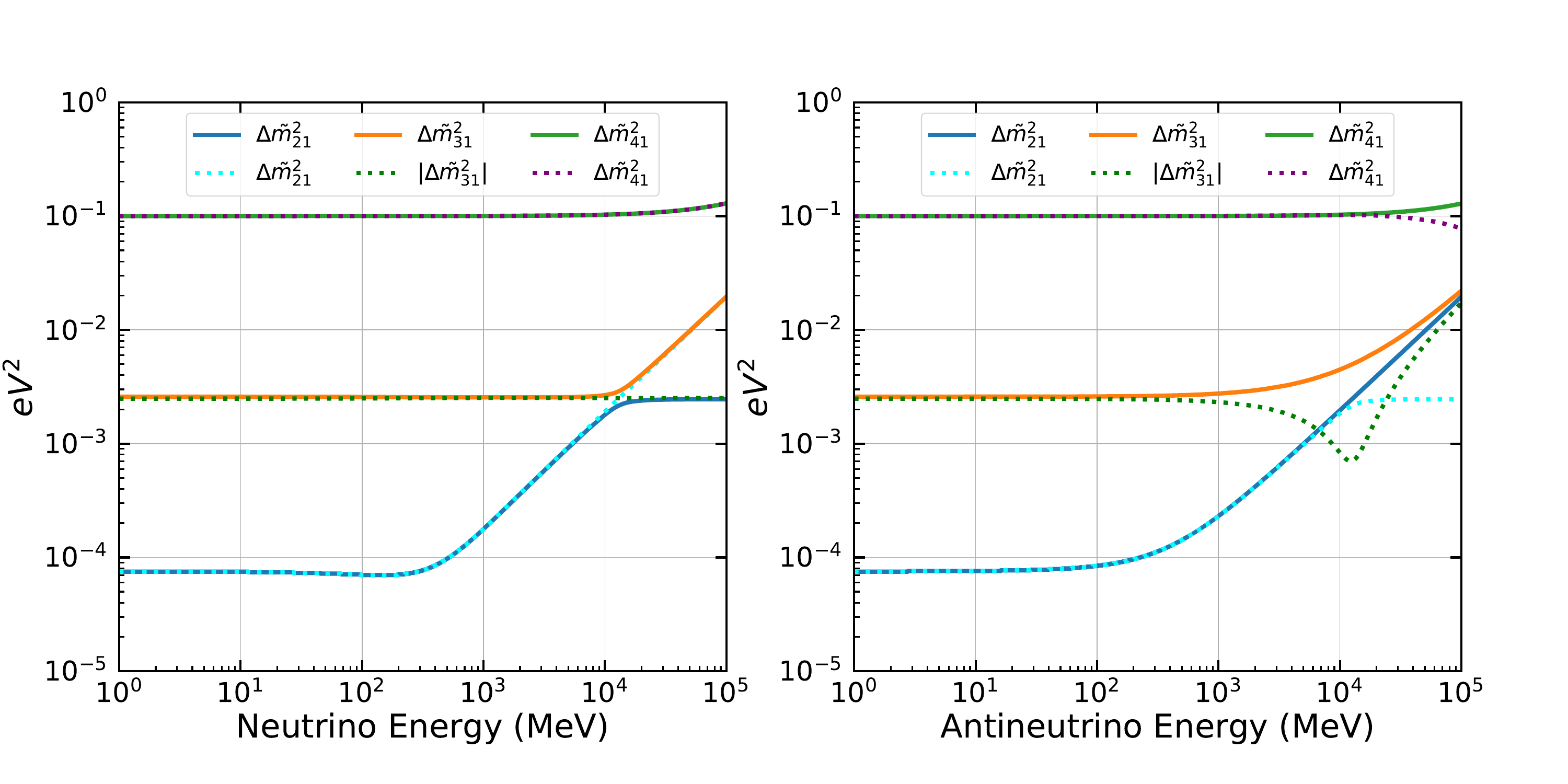}
\caption{\label{fig:2} The values of $\Delta\widetilde{m}_{i1}^2(i=2,3,4)$ with respect to neutrino energy, 
assuming $\Delta m_{41}^2=0.1$ eV$^2$, $\sin^2\theta_{14}=0.019$, $\sin^2\theta_{24}=0.015$, $\sin^2\theta_{34}=0$~\cite{cite:globalfit},
 $\delta_{13}=218^\circ$~\cite{cite:Neutrino2018} and $\delta_{24}=\delta_{34}=0^\circ$.
The solid and dashed lines are for NH and IH.}
\end{figure}

Up to now all the effective parameters in 4-flavor neutrino oscillation have been provided, and hence the neutrino oscillation probabilities can be easily calculated using eq.~\ref{eq:EffPro}. 
Since both CC and NC potentials in matter are proportional to neutrino energy, the values of those effective parameters in $\widetilde{U}$ 
and $\widetilde{m}_i^2\,(i=1,2,3,4)$ are also energy dependent, as shown in figure~\ref{fig:1}, \ref{fig:2}.  


\subsection{Discussion}
\label{sec:Analysis}
The effective matrix $\widetilde{U}$ in matter has introduced two effective mixing angles 
$\widetilde{\theta}_{12}$ and $\widetilde{\theta}_{13}$,
two effective CP phases $\widetilde{\delta}_{12}$ and $\widetilde{\delta}_{13}$, 
and effective mass-squared differences $\Delta \widetilde{m}_{ij}^2$, 
in which $\widetilde{\delta}_{12}$ is an additional parameter introduced from the Jacobi-like method. %none of which appear in the PMNS matrix $U$ in vacuum.
These effective parameters are clearly energy dependent, as shown in figure~\ref{fig:1}, \ref{fig:2}.

In figure~\ref{fig:1}, when $E<100$ MeV, $\widetilde{\theta}_{12}$ and $\widetilde{\theta}_{13}$ are very close to 
$\theta_{12}$ and $\theta_{13}$ values in vacuum. 
The value of $\widetilde{\theta}_{12}$ increases (decreases) rapidly up to the maximum $\frac{\pi}{2}$ 
( the minimum $0$) in the neutrino (antineutrino) energy range from $100$ MeV to $10$ GeV, leading to $\sin\widetilde{\theta}_{12}\rightarrow 1$ ($\sin\widetilde{\theta}_{12}\rightarrow 0$).
While $\widetilde{\theta}_{13}$ starts to change after $E>1$GeV. It can go up to $\frac{\pi}{2}$ assuming NH for neutrinos
and IH for antineutrinos when $E>100$ GeV; While it will go down to $0$ for the other two combinations.  
When $E<1$ GeV, both effective CP phases are close to their corresponding vacuum oscillation values ($\widetilde{\delta}_{12}\rightarrow 0$ and $\widetilde{\delta}_{13}\rightarrow\delta_{13}$). 
When energy increases above $1$ GeV, the influence of matter effect on $\widetilde{\delta}_{12}$ and $\widetilde{\delta}_{13}$ is not negligible. 

%When energy goes up, $\widetilde{\theta}_{13}$ is close to $\pi/2$ for neutrino-NH and antineutrino-IH cases. Then $\widetilde{\delta}_{13}$ will have a shift with $\pi$. It is caused by the limitation on $\widetilde{\delta}_{13}$ within $[0,\pi/2]$
%\footnote{$\widetilde{\theta}_{13}>\pi/2$ results in $\cos\widetilde{\theta}_{13}<0$. We set $\widetilde{\theta}_{13}$ within $[0,\pi/2]$, leading $\cos\widetilde{\theta}_{13}\equiv-\cos(\pi-\widetilde{\theta}_{13})\equiv\cos(\pi-\widetilde{\theta}_{13}) e^{i\pi}$. 
%And the phase $\pi$ will be added on $\widetilde{\delta}_{13}$.}.

In figure~\ref{fig:2}, the effect of matter also changes the values of effective neutrino mass-squared differences 
$\Delta \widetilde{m}_{ij}^{2}$.
When $E<100$ MeV, $\Delta \widetilde{m}_{21}^2$, $\Delta \widetilde{m}_{31}^2$ 
and $\Delta \widetilde{m}_{41}^2$ are close to their vacuum values.
$\Delta \widetilde{m}_{21}^2$ begins to vary when $E>100$ MeV, 
while $\Delta \widetilde{m}_{31}^2$ starts to change values after $E>1$ GeV. 
In the case of $\Delta m_{41}^2 = 0.1$ eV$^2$ with current sterile neutrino limits, $\Delta \widetilde{m}_{41}^2$ is insensitive to matter effect when $E<100$ GeV.  
As neutrino energy increases, 
matter effect shifts the values of effective $\Delta \widetilde{m}_{21}^2$ more than 
$\Delta \widetilde{m}_{31}^2$ and $\Delta \widetilde{m}_{41}^2$ when $E<100$ GeV. 
It should be noticed that $|\Delta\tilde{m}_{31}^2|$ has a dip structure around $10$ GeV for the antineutrino IH case. 
This feature also shows up in the 3-flavor neutrino case. 

In general, as shown in figure~\ref{fig:1} and \ref{fig:2}, matter effect is negligible on both $\Delta \widetilde{m}_{21}^2$ and 1-2 neutrino mixing 
when $A_{\mathrm{CC}}$ ($A_{\mathrm{NC}}$) $\ll \Delta m_{21}^2 \ll \Delta m_{31}^2$ (or equivalently $E\ll100$ MeV). %neutrino energy $E<100$ MeV. 
When energy increases, it is clear that the 1-2 neutrino mixing submatrix is affected 
from matter much more than other submatrices, as well as $\Delta \widetilde{m}_{21}^2$.
However, $\Delta \widetilde{m}_{31}^2$ and 1-3 mixing neutrino mixing still hold stable when $A_{\mathrm{CC}}$ ($A_{\mathrm{NC}}$) $\ll \Delta m_{31}^2$ (or equivalently $E\ll1$ GeV).
Furthermore, mixing between active and sterile neutrinos has little impact. 
From a mathematical point of view, 
in the function of rotation angles yielding 
$\tan\theta=\frac{2A}{\gamma-\alpha \pm \sqrt{(\gamma-\alpha)^2+4A^2}}$, 
$A$ is proportional to the values of $\theta_{14}$, $\theta_{24}$ and $\theta_{34}$, and $\gamma-\alpha$ is inversely proportional to $\Delta m_{41}^2$.
Hence, the smallness of those mixing angles and large $\Delta m_{41}^2(>0.1eV^2)$ can effectively suppress the values of the corresponding rotation angles to a negligible level in the submatrices. 
Therefore, after the rotations on 1-2 and 2-3 (1-3) submatrices of the neutrino Hamiltonian, 
 the effective Hamiltonian matrix is approximately diagonal.
%The above analysis is the reason why the effective sterile neutrino parameters 
%(i.e. $\widetilde{\theta}_{14}$, $\widetilde{\theta}_{24}$, $\widetilde{\theta}_{34}$, $\widetilde{\delta}_{24}$, $\widetilde{\delta}_{34}$, $\Delta \widetilde{m}_{41}^2$) 
%can be treated as the values in vacuum for our approximate analytical expressions.

What we have discussed is for the general feature of our derived oscillation formula. In some particular cases, the oscillation formula can be simplified:

\begin{itemize}
\item No CP violations ($\delta_{13}=\delta_{24}=\delta_{34}=0/\pi$)
\\* 
In such case, the neutrino mixing matrix is real and not necessary to introduce an extra phase $\widetilde{\delta}_{12}$ in eq.~\eqref{eq:EffU} for the matrix diagonalization. Thus, it falls back to the original Jacobi method. The neutrino oscillation forms are identical to the ones in vacuum with  $\widetilde{\theta}_{12}=\theta_{12}+\omega_1$, $\widetilde{\theta}_{13}=\theta_{13}+{\omega_2}$ and $\widetilde{\delta}_{13}=\delta_{13}$.

\item No active-sterile neutrino mixing ($\theta_{14}=\theta_{24}=\theta_{34}=\delta_{24}=\delta_{34}=0$)
\\*
The analytical approximations will reduce to 3-flavor neutrino oscillations.
\end{itemize}

\section{The accuracy of the approximations}
\label{sec:accuracy}

All the neutrino oscillation probabilities can be expressed with eq.~\eqref{eq:EffPro} based on
effective $\widetilde{U}$ and $\Delta\widetilde{m}_{ij}^2$ calculated in section~\ref{sec:Appro:EffUM}.
In this section, we first check the accuracy of these approximations. Then, we will highlight the accuracy of this work for two specific long-baseline accelerator neutrino experiments, T2HK and DUNE respectively.

\subsection{The general accuracy}

The accuracy of our approximations can be quantified with $\Delta P_{\nu_\alpha\rightarrow\nu_\beta}$, which is defined as the numerical difference between the approximations and exact solutions for neutrino  with $\alpha$ flavor type converting to $\beta$ type.
 
\begin{equation}
\Delta P_{\nu_\alpha\rightarrow\nu_\beta}=\left|P_{\nu_\alpha\rightarrow\nu_\beta}^{\rm Exact}-P_{\nu_\alpha\rightarrow\nu_\beta}^{\rm Approximate}\right| \,.
\end{equation}

\begin{figure}[htbp]
\centering
\includegraphics[width=0.9\textwidth]{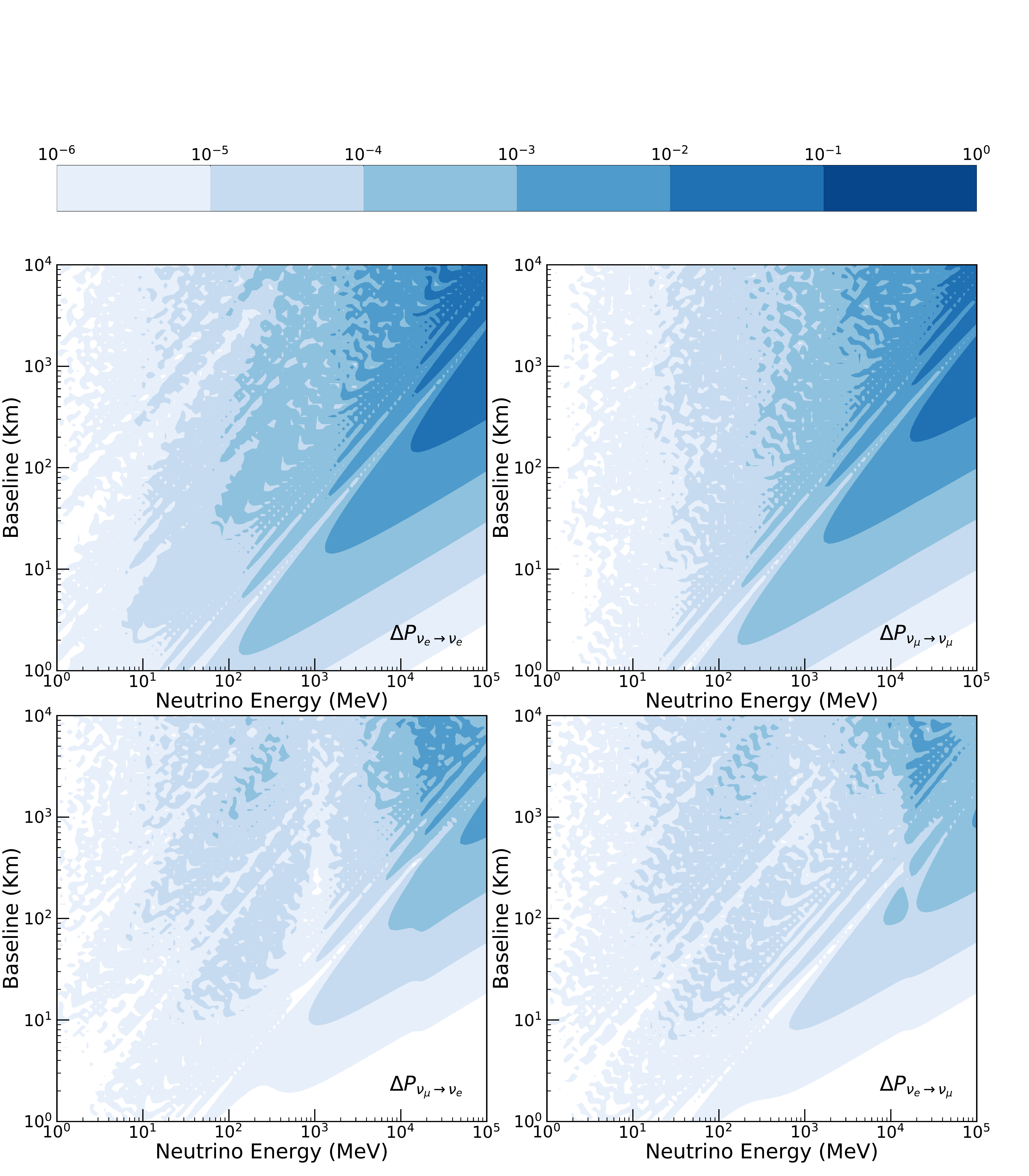}
\caption{\label{fig:3} 
The accuracies of approximations in different oscillation channels with $\Delta P_{\nu_\alpha\rightarrow\nu_\beta}$.}
\end{figure}

For checking the validity of our approximations, Figure~\ref{fig:3} presents the results, as a function of neutrino energy and travel baseline, on four major neutrino oscillation channels, including $\nu_{e} \rightarrow \nu_{e}$, $\nu_{\mu} \rightarrow \nu_{\mu}$ disappearance, and $\nu_{\mu} \rightarrow \nu_{e}$ and $\nu_{e} \rightarrow \nu_{\mu}$ appearance. The input values for the oscillation parameters are listed in table~\ref{table:i}.  Here we conservatively assume the unknown sterile neutrino associated mixing angles to be as large as $20^\circ$ and unknown phases to be maximal, $90^\circ$. 

\begin{table}[htbp]
\centering
\begin{tabular}{|c|c|c|c|c|}
\hline
\multicolumn{3}{|c|} {active} & \multicolumn{2}{|c|}{sterile} \\
\hline
& NH & IH & \multicolumn{2}{|c|} {} \\
\hline
${\sin^2\theta_{12}}$ &\multicolumn{2}{|c|} {0.307} &${\theta_{14}}$ & $20^\circ$ \\
\hline
${\sin^2\theta_{13}}$ &\multicolumn{2}{|c|} {0.0212} &${\theta_{24}}$ & $20^\circ$ \\
\hline
${\sin^2\theta_{23}}$ & 0.417 & 0.421 &${\theta_{34}}$ & $20^\circ$ \\
\hline
${\Delta m^2_{21}}$ & \multicolumn{2}{|c|} {$7.53\times 10^{-5}$ eV$^2$} &${\Delta m^2_{41}}$ &0.1 eV$^2$\\
\hline
${\Delta m^2_{32}}$ & $2.51 \times 10^{-3}$ eV$^2$ & $-2.56 \times 10^{-3}$ eV$^2$ &${\delta_{24}}$ & $\pi/2$\\
\hline
$\delta_{13}$ & \multicolumn{2}{|c|} {$\pi/2$} & $\delta_{34}$ & $\pi/2$\\
\hline
\end{tabular}
\caption{\label{table:i} Input values for the oscillation parameters. The ones associated with active neutrino except $\delta_{13}$ are taken from the recent results~\cite{cite:Neutrino2018}. The sterile neutrino values are conservatively using relatively large values.}
\end{table} 

As shown in Figure~\ref{fig:3}, when neutrino energy below $20$ GeV, the accuracy is better than $10^{-3}$ and $10^{-4}$ for neutrino disappearance and appearance channels respectively. The accuracy will get one or two orders better when neutrino energy is smaller. Such numerical accuracy of our approximations is pretty good for most of the long baseline neutrino oscillation experiments, since the difference is about one order smaller than the oscillation probabilities. The inaccuracy of our approach is mainly caused by the limited rotation iterations we applied. One approach to further improve the accuracy is to apply some corrections based on perturbative method after matrix rotation, as shown in Ref~\cite{cite:3+1MSW}. Of course, this will add the complexity of the analytical expression. Our current approximations give a good balance between the numerical accuracy and simplicity of the approximation expressions. To be noted, the exact accuracy of the approximations depends on the exact neutrino mixing parameter inputs. Once the sterile neutrino mixing angles to be smaller than $20^{\circ}$ and $\Delta m^{2}_{41}$ be larger than $0.1$ eV$^{2}$, the accuracy of our expressions will become even better than the current evaluation.     
 \begin{figure}[htb]
\centering
\includegraphics[width=0.9\textwidth]{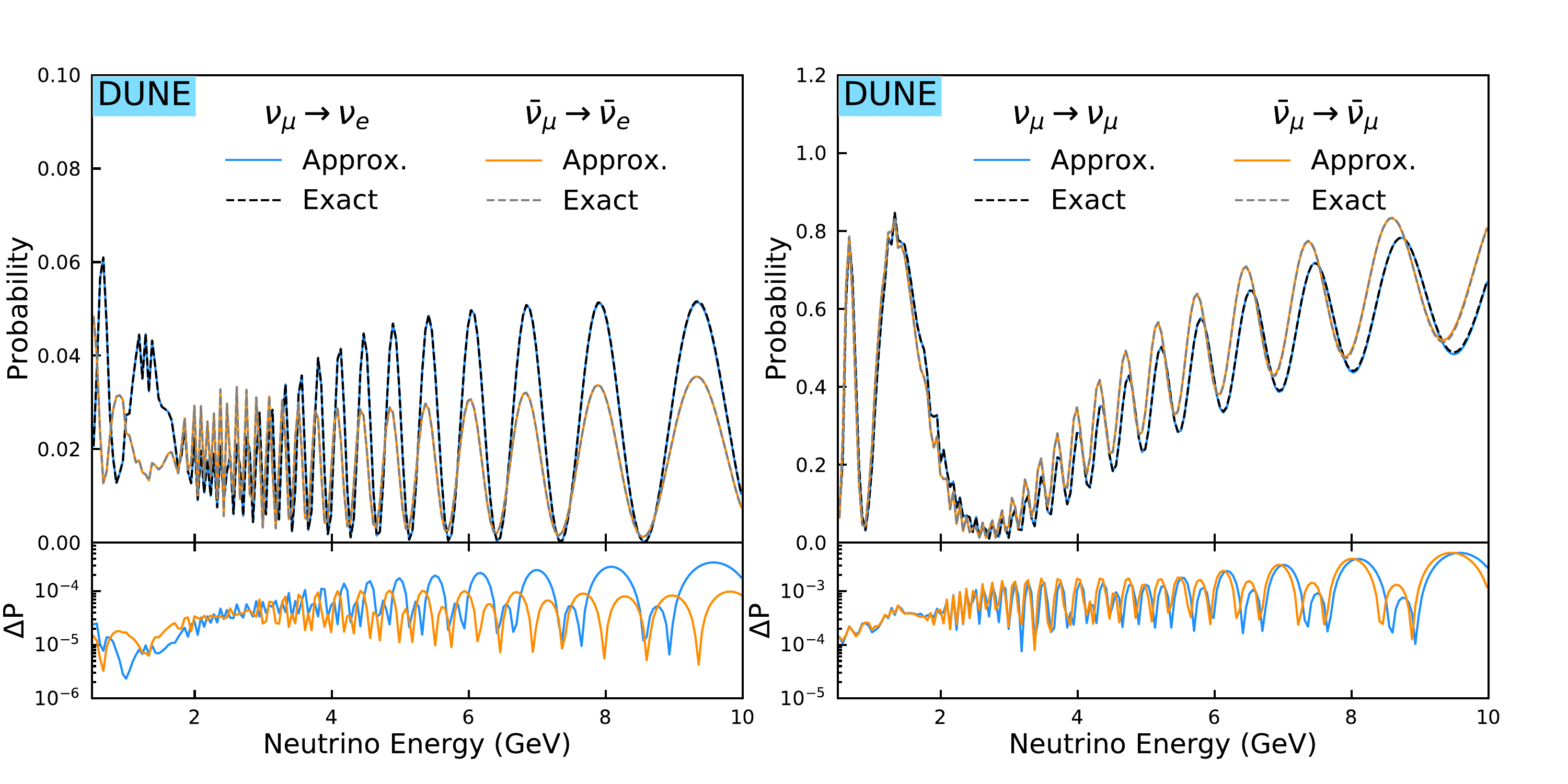}
\includegraphics[width=0.9\textwidth]{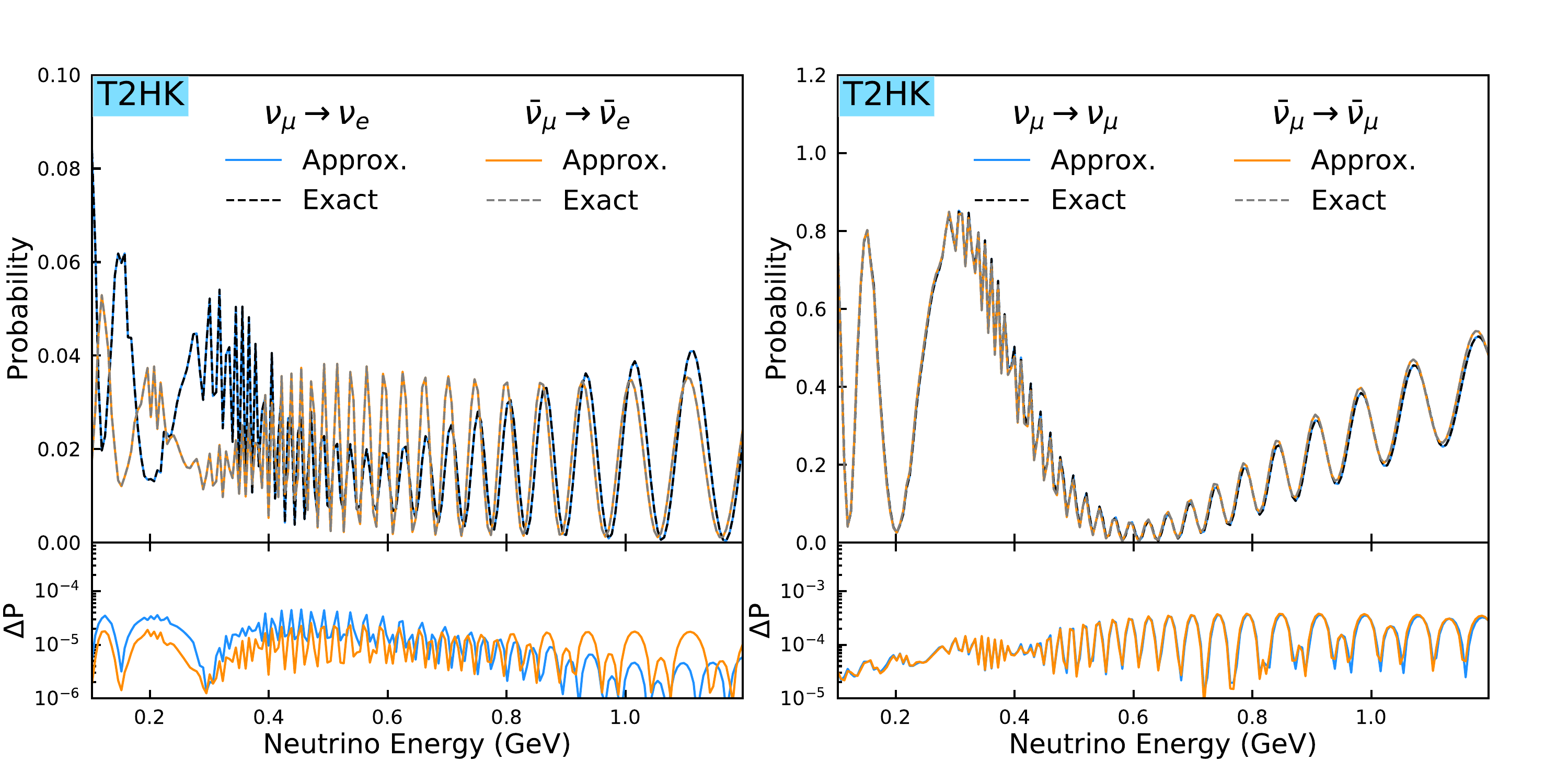}
\caption{\label{fig:4}  The left plots are appearance channels and 
the right plots are disappearance channels in the case of NH. 
The appendant plots are accuracies of their corresponding channels.}
\end{figure}
\subsection{T2HK and DUNE}
As shown in eq.~\eqref{eq:NeuH} and \eqref{eq:A}, matter effect is proportional to neutrino beam energy and its propagation distance. 
Hence matter effect can significantly modify neutrino oscillation features for long-baseline neutrino oscillation experiments, 
such as T2HK and DUNE.
Those experiments have relative high energy beams at $\sim$GeV and baselines of hundreds and thousands of kilometers respectively.
Here we would like to use them as examples to demonstrate 3+1-neutrino oscillation 
and check the accuracy of our approximations. 

Deep Underground Neutrino Experiment (DUNE) is the next generation on-axis long-baseline accelerator neutrino experiment.
It is proposed to use Liquid Argon (LAr) detectors located deep underground $1300$ km away from the beam source. 
Its main physics goals are to solve three challenging issues in the neutrino sector, 
neutrino mass hierarchy, CP asymmetry and the octant of $\theta_{23}$. 
It can look for electron and tau neutrino (anti-neutrino) appearance and muon neutrino (antineutrino) disappearance channels 
from both neutrino and antineutrino beam modes. 
%Here, we treat the earth density $\rho$ as a constant $3$ g/cm$^3$.

\label{sec:example}  
\begin{figure}[htb]
\centering
\includegraphics[width=\textwidth]{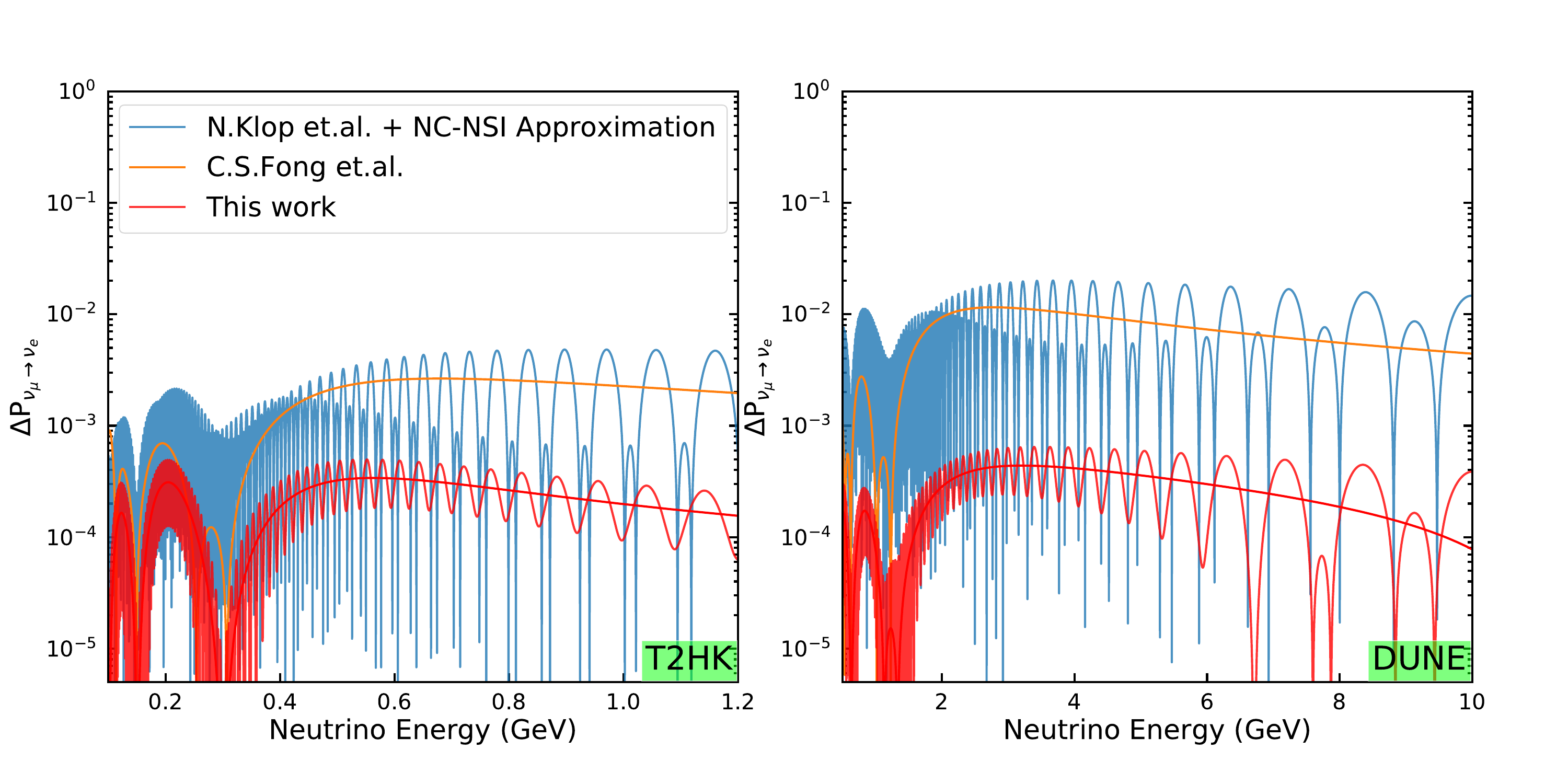}
\caption{\label{fig:compare}  The comparison of $P_{\bar{\nu}_\mu\rightarrow\bar{\nu}_e}$ for DUNE and T2HK in the case of NH with the the values in table~\ref{table:i} except the phases. 
We adopt $\delta_{13}=90^\circ$ and $\delta_{14}=90^\circ$ due to the convention N. Klop et. al. \cite{cite:Klop} use.
Similar to the examples shown in their paper, we combine it with a Neutral Current NSI solution \cite{cite:NCNSI} to produce the oscillation  probability.}
\end{figure} 
 
Tokai-to-Hyper-Kamiokande (T2HK) is a proposed long-baseline experiment, which has a primary objective of measuring CP asymmetry.
The far detector is $295$ km away and $2.5^\circ$ off-axis from the J-PARC beam in Japan using water Cherenkov detector. 

Suppose the existence of the sterile neutrinos with relatively large mass-squared difference $\Delta m_{41}^2 =0.1$eV$^2$, 
the high frequency oscillation feature is clearly shown in the muon neutrino disappearance and electron neutrino appearance modes in figure~\ref{fig:4}.
Given matter effect and CP-violation phases, the electron neutrino and antineutrino appearance probabilities 
$P_{\nu_\mu\rightarrow\nu_e}$ and $P_{\bar{\nu}_\mu\rightarrow\bar{\nu}_e}$ are very different.  
Compared with numerical calculation, the accuracy of the analytical approximations can reach $10^{-5}$ in the case of table~\ref{table:i}.
for neutrinos and antineutrinos  respectively for appearance mode.
For disappearance mode, the accuracies of $P_{\nu_\mu\rightarrow\nu_\mu}$ and $P_{\bar{\nu}_\mu\rightarrow\bar{\nu}_\mu}$ can reach $10^{-4}$ in the case of table~\ref{table:i}.
%As shown using the above experiments, our approximations can extend to the other long baseline accelerator experiments with accuracy under current sterile neutrino limits. 
%To be noted, our approximations are more general. It can not only be applied to long baseline experiments, 
%but also short and medium baseline neutrino experiments, such as JUNO~\cite{cite:junomattereffect}, although their matter effects are usually quite small and can be neglected. 

In figure~\ref{fig:compare}, we compare the accuracy of this work with two previous ones for the T2HK and DUNE experiments.
Our work clearly shows about an order of better accuracy compared with the other two, especially for the DUNE experiment.  
For a higher accuracy of the approximation, we can continue making a perturbation correction afterwards on the effective neutrino mixing and mass-squared differences as the way \cite{cite:3+1MSW} adopts.
However, given the accuracy of this work is already pretty good for the current and near future experiments,  we don't think it is really necessary.

\section{Summary}
\label{sec:Summary}
The search for light sterile neutrinos is an area of great interest in the neutrino field. Many long baseline neutrino experiments are very actively searching for light sterile neutrinos in various mass regions. Both charged-current and neutral-current induced matter effects are quite important for those experiments. Analytical approximations of neutrino oscillation are more preferred in experimental neutrino research because of its considerable time-saving and helpful for understanding neutrino oscillation features.

In this manuscript we introduced a Jacobi-like method to derive simplified analytical expressions with good accuracy for neutrino oscillation in matter. 
The compact expressions of the effective mixing matrix $\widetilde{U}$ and the effective
mass-squared differences $\Delta\widetilde{m}_{ij}^2(i,j=1,2,3,4)$ are presented.
The accuracy of this work is sufficient for the majority of long baseline neutrino experiments.

In addition, the Jacobi-like method is a general method to diagonalize complex Hermitian matrices. It also can be extended into other physics topics, such as 3 (active) + N (sterile)-neutrino mixing, Neutrino Non-Standard Interactions, etc.

\section{Acknowledgments}
We would like to extend our thanks for valuable discussions with Yu-Feng Li and help with editing from Neill Raper.  Jiajie Ling acknowledges the support from National Key R\&D program of China under Grant No. 2018YFA0404103 and National Natural Science Foundation of China under Grant No. 11775315. F. Xu is supported partially by NSFC under Grant No. 11605076, as well as the FRFCU (Fundamental Research Funds for the Central Universities in China) under the Grant No. 21616309.

\bibliography{references}

\appendix
\section{The parameterization of mixing matrix in vacuum}
\label{sec:appendixA}
In the 3 + 1 framework, neutrino mixing can be written as a $4\times 4$ matrix~\eqref{eq:4mixing}. 6 rotation angles with 3 Dirac phases\footnote{In this table, we adopt $\delta_{24}$ and $\delta_{34}$ as the sterile phases. It is equivalent to using $\delta_{14}$ and $\delta_{24}$, or $\delta_{14}$ and $\delta_{34}$ for additional CP phases.} 
are found in this matrix. All the elements of the mixing matrix are listed in table~\ref{table:1}.
Indeed, if we set the angles and Dirac phases introduced by sterile neutrinos to 0, the $4\times4$ matrix will reduce to 3-flavor neutrino mixing.

\begin{table}[htbp]
 %\label{table:1}
 \renewcommand{\multirowsetup}{\centering}
 \centering
\begin{tabular}{|c|c|c|}
\hline
$\alpha$& $U_{\alpha i}$ &  -\\
\hline
\multirow{4}*{$e$}&$U_{e1}$&
$c_{12}c_{13}c_{14}$
\\
\cline{2-3}
&$U_{e2}$&
$c_{13}c_{14}s_{12}$
\\
\cline{2-3}
&$U_{e3}$&
$c_{14}s_{13}e^{-i\delta_{13}}$
\\
\cline{2-3}
&$U_{e4}$&
$s_{14}$
\\
\hline
\multirow{4}*{$\mu$}&$U_{\mu1}$&
$-s_{12}c_{23}c_{24}-c_{12}(s_{13}c_{24}s_{23}e^{i\delta_{13}}+c_{13}s_{14}s_{24}e^{-i\delta_{24}})$
\\
\cline{2-3}
&$U_{\mu2}$&
$c_{12}c_{23}c_{24}-s_{12}(s_{13}c_{24}s_{23}e^{i\delta_{13}}+c_{13}s_{14}s_{24}e^{-i\delta_{24}})$
\\
\cline{2-3}
&$U_{\mu3}$&
$c_{13}c_{24}s_{23}-s_{13}s_{14}s_{24}e^{-i\delta_{13}}e^{-i\delta_{24}}$
\\
\cline{2-3}
&$U_{\mu4}$&
$c_{14}s_{24}e^{-i\delta_{24}}$
\\
\hline
\multirow{4}*{$\tau$}&$U_{\tau1}$&
$\begin{aligned}  
c_{12}[s_{13}(s_{23}s_{24}s_{34}e^{i\delta_{24}}e^{-i\delta_{34}}-&c_{23}c_{34})e^{i\delta_{13}}-c_{13}c_{24}s_{14}s_{34}e^{-i\delta_{34}}] \\
+s_{12}(c_{34}s_{23}+c_{23}&s_{24}s_{34}e^{i\delta_{24}}e^{-i\delta_{34}})
\end{aligned}$
\\
\cline{2-3}
&$U_{\tau2}$&
$\begin{aligned}  s_{12}[s_{13} (s_{23}s_{24}s_{34}e^{i\delta_{24}}e^{-i\delta_{34}}-&c_{23}c_{34})e^{i\delta_{13}} -c_{13}c_{24}s_{14}s_{34}e^{-i\delta_{34}}]
\\
-c_{12}(c_{34}s_{23}+c_{23}&s_{24}s_{34}e^{i\delta_{24}}e^{-i\delta_{34}})
\end{aligned}$\\
\cline{2-3}
&$U_{\tau3}$&
$c_{13}(c_{23}c_{34}-s_{23}s_{24}s_{34}e^{i\delta_{24}}e^{-i\delta_{34}})-s_{13}c_{24}s_{14}s_{34}e^{-i\delta_{13}}e^{-i\delta_{34}}$
\\
\cline{2-3}
&$U_{\tau4}$&
$c_{14}c_{24}s_{34}e^{-i\delta_{34}}$
\\
\hline
\multirow{4}*{$s$}&$U_{s1}$&
$\begin{aligned}
c_{12}[s_{13}(c_{34}s_{23}s_{24}e^{i\delta_{24}}+c_{23}s_{34}&e^{i\delta_{34}})e^{i\delta_{13}}-c_{13}c_{24}c_{34}s_{14}]
\\
+
s_{12}(c_{23}c_{34}s_{24}e^{i\delta_{24}}-&s_{23}s_{34}e^{i\delta_{34}})
\end{aligned}$
\\
\cline{2-3}
&$U_{s2}$&
$
\begin{aligned}
s_{12}[s_{13}
(c_{34}s_{23}s_{24}e^{i\delta_{24}}+c_{23}s_{34}&e^{i\delta_{34}})e^{i\delta_{13}}-c_{13}c_{24}c_{34}s_{14}]
\\
+c_{12}(s_{23}s_{34}e^{i\delta_{34}}-&c_{23}c_{34}s_{24}e^{i\delta_{24}})
\end{aligned}$
\\
\cline{2-3}
&$U_{s3}$&
$-c_{13}(c_{34}s_{23}s_{24}e^{i\delta_{24}}+c_{23}s_{34}e^{i\delta_{34}})-c_{24}c_{34}s_{14}e^{-i\delta_{13}}$
\\
\cline{2-3}
&$U_{s4}$&
$c_{14}c_{24}c_{34}$
\\
\hline
\end{tabular}
\caption{\label{table:1} The elements of the 4-flavor mixing matrix.}
\end{table}

 \section{Jacobi-like method}
\label{sec:appendixB}
 Matter effect for the 3+1 framework is more difficult to calculate than for the 3 neutrino framework because additional parameters are 
 involved in neutrino Hamiltonian, which is a $4\times4$ complex Hermitian matrix and difficult to diagonalize.
 In this work, we adopt a rotation technique known as the Jacobi-like method to solve the diagonalization for complex Hermitian matrices. In the following, we provide all technical details for the neutrino and antineutrino cases separately. 
 \subsection{Neutrino case}
 \label{sec:NeutrinoCase}
 In this subsection, we show how to diagonalize the effective Hamiltonian with matter effect and simplify the expressions of the effective mixing and mass-squared differences for the neutrino case.
\subsubsection{Diagonalization process}
\label{sec:NeutrinoCaseRotations}
In the case of neutrinos, we find the absolute values of elements $H_{12}$ and $H_{21}$
in eq.~\eqref{eq:H} are the relatively largest off-diagonal ones because of the smallness of $\Delta m_{21}^2$.
Hence, we should diagonalize the 1-2 submatrix first. 

\paragraph{First rotation:}
The rotation matrix can be written as
\begin{equation}
\label{eq:NeuRoStep1}
R^1=R_{12}({\omega_1},\phi_1)\equiv \begin{bmatrix}
c_{{\omega_1}}&s_{{\omega_1}}e^{-i\phi_1}&0&0\\
-s_{{\omega_1}}e^{i\phi_1}&c_{{\omega_1}}&0&0\\
0&0&1&0\\
0&0&0&1
\end{bmatrix}\,. \qquad (c_{{\omega_1}}=\cos{\omega_1} \,,s_{{\omega_1}}=\sin{\omega_1})
\end{equation}
We utilize the Jacobi-like method to derive ${\omega_1}$ yielding
\begin{equation}
\label{eq:NeutTan1}
\tan{\omega_1}=
\frac{2A_{{\omega_1}}}{(H_{22}-H_{11})+
\sqrt{(H_{22}-H_{11})^2+4A_{{\omega_1}}^2}}
\,,\qquad(0< {\omega_1} <\frac{\pi}{2}-{\theta}_{12})
\end{equation}
with $A_{{\omega_1}}=|H_{12}|$ and $\phi_1=\mathrm{Arg}(\mathrm{sign}(A_{\omega_1})H_{12}^{*})$. Here $A_{{\omega_1}}$ is the amplitude of $H_{12}$.
After the rotation by $R_{12}({\omega_1},\phi_1)$, we rewrite the Hamiltonian in the new representation as
\begin{equation}
\label{eq:NeuHPrime}
H^\prime = R_{12}^\dag({\omega_1},\phi_1) H R_{12}({\omega_1},\phi_1)
=
\frac{1}{2E}
\begin{bmatrix}
\lambda_{-} &0 &H_{13}^{\prime} & H_{14}^{\prime } \\
0 & \lambda_{+} & H_{23}^{\prime } & H_{24}^{\prime} \\
H_{31}^{\prime } & H_{32}^{\prime } & H_{33} & H_{34} \\
H_{41}^{\prime } & H_{42}^{\prime } & H_{43}& H_{44}
\end{bmatrix}\,,
\end{equation}
with the eigenvalues of the 1-2 submatrix 
\begin{equation}\label{eq:NeuHPrimeEle}
\lambda_-=\frac{H_{11}+H_{22}\tan^2{\omega_1}-2A_{\omega_1}\tan{\omega_1}}
{1+\tan^2{\omega_1}} \,,
\lambda_+=\frac{H_{11}\tan^2{\omega_1}+H_{22}+2A_{\omega_1}\tan{\omega_1}}
{1+\tan^2{\omega_1}} \,.
\end{equation}

The corresponding off-diagonal terms in $H^\prime$ are
\begin{equation}
H_{13}^\prime = H_{31}^{\prime *}=c_{\omega_1}H_{13}-s_{\omega_1} H_{23}e^{-i\phi_1} \,,
\end{equation}
\begin{equation}
H_{23}^\prime=H_{32}^{\prime *} = c_{\omega_1}H_{23} + s_{\omega_1} H_{13}e^{i\phi_1} \,,
\end{equation}
\begin{equation}
H_{14}^\prime = H_{41}^{\prime *} = c_{\omega_1}H_{14} - s_{\omega_1} H_{24}e^{-i\phi_1} \,,
\end{equation}
\begin{equation}
H_{24}^\prime= H_{42}^{\prime *}= c_{\omega_1}H_{24} + s_{\omega_1} H_{14}e^{i\phi_1} \,.
\end{equation}

%$H_{23}^\prime=\frac{H_{13}\tan{\omega_1} e^{i\phi_1}+H_{23}}{\sqrt{1+\tan^2{\omega_1}}}$ is useful term for the further calculations.

After first rotation, we can diagonalize the 2-3 submatrix afterward since it has a relatively big off-diagonal element and is useful to simplify
the effective matrix $U$ through eq.~\eqref{eq:NeuExch2}. 

\paragraph{Second rotation:}
The second rotation matrix yields
\begin{equation}
\label{eq:NeuRoStep2}
R^2=R_{23}({\omega_2},{\phi_2}) \equiv \begin{bmatrix}
1&0&0&0\\
0&c_{{\omega_2}}&s_{{\omega_2}}e^{-i{\phi_2}}&0\\
0&-s_{{\omega_2}}e^{i{\phi_2}}&c_{{\omega_2}}&0\\
0&0&0&1
\end{bmatrix}\,.\qquad (c_{{\omega_2}}=\cos{\omega_2},s_{{\omega_2}}=\sin{\omega_2})
\end{equation}
The diagonal angle ${\omega_2}$ is compatible with
\begin{equation}
\label{eq:NeuTan2}
\tan{\omega_2}
=
\frac{2A_{\omega_2}}{(H_{33}-\lambda_+)\pm
\sqrt{(H_{33}-\lambda_+)^2+4A_{\omega_2}^2}} \,,
\end{equation}
where the upper sign is for NH $ (0 < {\omega_2} < \frac{\pi}{2}-\theta_{13}) $,
and the lower sign is for IH $( 0 > {\omega_2} > -\theta_{13}) $.
$A_{{\omega_2}}$ is the amplitude of $H_{23}$ in eq.~\eqref{eq:NeuHPrime} yielding $A_{{\omega_2}}=|H_{23}^\prime|$.
The additional complex factor is given by $\phi_2=\mathrm{Arg}(\mathrm{sign}(A_{\omega_2})H_{23}^{\prime*})$.
After two rotations, we obtain
\begin{equation}
\label{eq:NeuHPrimePrime}
H^{\prime\prime}= R_{23}^\dag({\omega_2},{\phi_2}) H^{\prime} R_{23}({\omega_2},{\phi_2}) =
\frac{1}{2E} \begin{bmatrix}
\lambda_{-} &H_{12}^{\prime\prime}&H_{13}^{\prime\prime}&H_{14}^{\prime}\\
H_{21}^{\prime\prime}&\lambda_-^{\prime}&0&H_{24}^{\prime\prime}\\
H_{31}^{\prime\prime}&0&\lambda_{+}^{\prime}&H_{34}^{\prime\prime}\\
H_{41}^{\prime}&H_{42}^{\prime\prime}&H_{43}^{\prime\prime}&H_{44}
\end{bmatrix} \,,
\end{equation}
where the diagonal terms $\lambda_-^\prime$ and $\lambda_+^\prime$ obey
\begin{equation}
\label{eq:NeuLambPrime}
\lambda_-^\prime=\frac{\lambda_+ +H_{33}\tan^2{\omega_2}-2A_{\omega_2}\tan{\omega_2}}
{1+\tan^2{\omega_2}} \,,\quad
\lambda_+^\prime=\frac{\lambda_+\tan^2{\omega_2}+H_{33}+2A_{\omega_2}\tan{\omega_2}}
{1+\tan^2{\omega_2}} \,.
\end{equation}

The corresponding off-diagonal elements obey
\begin{equation}
H^{\prime\prime}_{12}=H^{\prime\prime*}_{21}= - s_{\omega 2} H_{13}^\prime e^{i\phi_2} \,,
\end{equation}
\begin{equation}
H^{\prime\prime}_{13}=H^{\prime\prime*}_{31}=c_{\omega 2} H_{13}^\prime  \,,
\end{equation}
\begin{equation}
H^{\prime\prime}_{24}=H^{\prime\prime*}_{42}= c_{\omega 2} H_{24}^\prime - s_{\omega 2} H_{34} e^{-i\phi_2} \,,
\end{equation}
\begin{equation}
H^{\prime\prime}_{34}=H^{\prime\prime*}_{43}= c_{\omega 2} H_{34} + s_{\omega 2} H_{24}^\prime e^{i\phi_2} \,.
\end{equation}

\begin{figure}[htbp]
\centering
\includegraphics[width=0.6\textwidth]{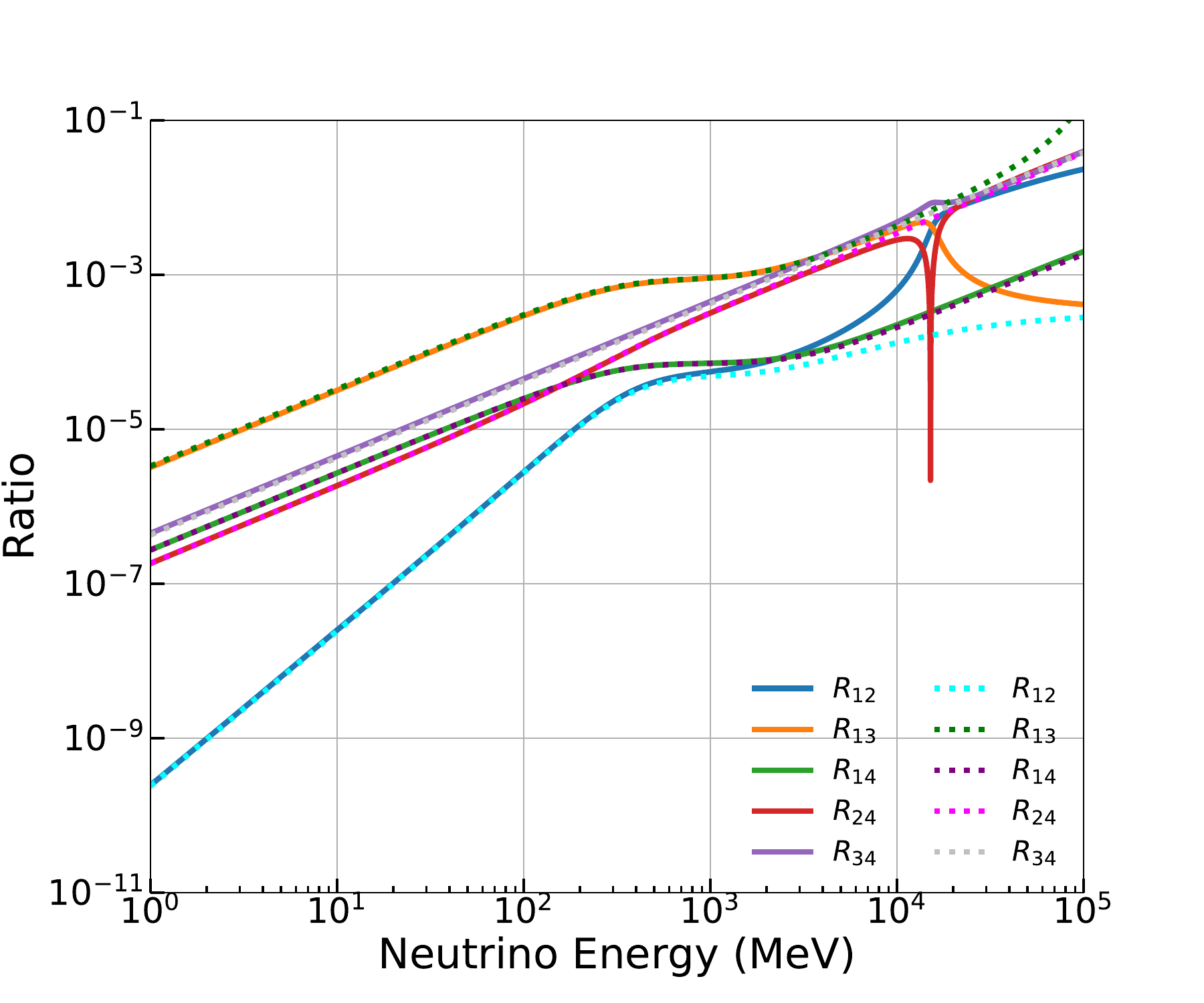}
\caption{\label{fig:5}  The values of the off-diagonal elements in $H^{\prime\prime}$. The solid and dashed lines are for NH and IH. The input oscillation parameters are listed in Table~\ref{table:i}.}
\end{figure}

After two continuous rotations, the matrix is approximately diagonalized. We can evaluate the matrix diagonalization using the ratio defined as
\begin{equation}
R_{ij}\equiv\left|\frac{H^{\prime\prime}_{ij}}{H^{\prime\prime}_{jj}-H^{\prime\prime}_{ii}}\right| \,(i<j), 
\end{equation}
which compares the relative size of the off-diagonal elements in $H^{\prime\prime}$ with the difference between the two corresponding diagonal terms.
Figure~\ref{fig:5} shows the ratios for various sub-matrices with respect to the neutrino energy. All of them are quite smaller than $0.1$, which means the corresponding rotation angles on $i-j$ submatrix of $H^{\prime\prime}$ will be much smaller than $5^\circ$ based on section~\ref{sec:Jacobi}.
Thus $H^{\prime\prime}$ matrix can be treated as an approximately diagonalized matrix in our application range. 

After diagonalization, the effective neutrino mixing matrix and mass-squared differences are given by
\begin{subequations}\label{eq:EffNeuDU}
\begin{gather}
\label{eq:EffNeuDU:1}
\widetilde{U}\approx UR_{12}(\omega_1,\phi_1)R_{23}({\omega_2},{\phi_2})
=\underbrace{R_{34}R_{24}R_{14}R_{23}R_{13}R_{12}}_{U}
R_{12}(\omega_1,\phi_1)R_{23}({\omega_2},{\phi_2}) \,,
\\
\label{eq:EffNeuDU:2}
\Delta \widetilde{m}_{21}^2\approx\lambda_{-}^{\prime}-\lambda_- \,,\quad
\Delta \widetilde{m}_{31}^2\approx\lambda_{+}^{\prime}-\lambda_- \,,\quad
\Delta \widetilde{m}_{41}^2\approx H_{44}-\lambda_- \,.
\end{gather}
\end{subequations}

If $\delta_{13}=\delta_{24}=\delta_{34}=0$, $\phi_1$ and ${\phi_2}$ will be 0.
For the sake of beauty in mathematical form (It is convenient for $\widetilde{U}$ to completely have the same form with $U$.)  
as well as a good understanding of matter effect on oscillation parameters, we continue to simplify
the effective mixing matrix $\widetilde{U}$ below.
 \subsubsection{Simplification}
 \label{sec:NeutrinoCaseAbsorpation}
Firstly, we find there are two 1-2 submatrix rotations $R_{12}(\theta_{12},0)$ and $R_{12}(\omega_1,\phi_1)$
next to each other in eq.~\eqref{eq:EffNeuDU:2}. Therefore we combine them into one submatrix.
The processes are given by
\begin{equation}
\label{eq:EffNeuR12}
\begin{split}
&R_{12}(\theta_{12},0)R_{12}(\omega_1,\phi_1)\\
=&
\begin{bmatrix}
c_{12}&s_{12}&0&0\\
-s_{12}&c_{12}&0&0\\
0&0&1&0\\
0&0&0&1
\end{bmatrix}
\begin{bmatrix}
c_{\omega_1}&s_{\omega_1}e^{-i\phi_1}&0&0\\
-s_{\omega_1}e^{i\phi_1}&c_{\omega_1}&0&0\\
0&0&1&0\\
0&0&0&1
\end{bmatrix}
\\=&
\begin{bmatrix}
\widetilde{c}_{12}&\widetilde{s}_{12}e^{-i\widetilde{\delta}_{12}}&0&0\\
-\widetilde{s}_{12}e^{i\widetilde{\delta}_{12}}&\widetilde{s}_{12}&0&0\\
0&0&1&0\\
0&0&0&1
\end{bmatrix}
\begin{bmatrix}
e^{i\Theta_{12}}&0&0&0\\
0&e^{-i\Theta_{12}}&0&0\\
0&0&1&0\\
0&0&0&1
\end{bmatrix}
\\=&
R_{12}(\widetilde{\theta}_{12},\widetilde{\delta}_{12})D_{12}(e^{i\Theta_{12}},e^{-i\Theta_{12}},1,1)
\\=&
\widetilde{R}_{12}
\end{split} \,.
\end{equation}
Then we obtain
\begin{equation}
\widetilde{U}\approx  R_{34}R_{24}R_{14}R_{23}R_{13}\widetilde{R}_{12}R_{23}({\omega_2},{\phi_2}) \,.
\end{equation}
We find there is a new $\widetilde{R}_{12}$ consisting of a rotation matrix with
$\widetilde{\theta}_{12}$ and $\widetilde{\delta}_{12}$ and one
diagonal unitary matrix with a phase $\Theta_{12}$.
Those new items yield
\begin{subequations}\label{eq:NeuNewPar12}
\begin{gather}
\label{eq:NeuNewPar12:1}
\widetilde{s}_{12}=
\sin\widetilde{\theta}_{12}
 = \frac{|c_{12}\tan\omega_1 e^{i\phi_1}+s_{12}|}{\sqrt{1+\tan^2\omega_1}} \,,
\quad
\widetilde{c}_{12}=
\cos\widetilde{\theta}_{12}
 = \frac{|c_{12}-s_{12}\tan\omega_1 e^{i\phi_1}|}{\sqrt{1+\tan^2\omega_1}} \,,
\\
\label{eq:NeuNewPar12:2}
e^{i\widetilde{\delta}_{12}}
=
\frac{(c_{12}\tan\omega_1 e^{i\phi_1}+s_{12})(c_{12}-s_{12}\tan \omega_1 e^{-i\phi_1})}
{\cos \widetilde{\theta}_{12} \sin \widetilde{\theta}_{12} (1+\tan^2 \omega_1 )} \,,
\\
\label{eq:NeuNewPar12:3}
e^{i\Theta_{12}}=\frac{c_{12}-s_{12}\tan\omega_1 e^{\phi_1}}
{\cos\widetilde{\theta}_{12}\sqrt{1+\tan^2\omega_1}} \,.
\end{gather}
\end{subequations}
Here, we set the same limit on $\widetilde{\theta}_{12}$ in $[0,\frac{\pi}{2}]$ with $\theta_{12}$. 
%The energy dependent $\widetilde{\theta}_{12}$ is shown in figure~\ref{fig:1}.
%When neutrino energy is small, matter effects are slight, leading to $\omega_2\rightarrow 0$.
%Hence, we get $\widetilde{R}_{12} R_{23}({\omega_2},{\phi_2})\approx R_{13}({\omega_2},{\phi_2})\widetilde{R}_{12}$. 
%When $E$ goes up , we find $\sin\widetilde{\theta}_{12} \rightarrow 1$,
%leading to
%\begin{equation}\label{eq:NeuExch}
% \widetilde{R}_{12} \approx
% \begin{bmatrix}
%  0&1&0&0\\
%  -1&0&0&0\\
%  0&0&1&0\\
%  0&0&0&1
% \end{bmatrix} \,,
%\end{equation}
%which can be used in the following operation
%\begin{equation}\label{eq:NeuExch1}
% \begin{bmatrix}
%  0&1&0&0\\
%  -1&0&0&0\\
%  0&0&1&0\\
%  0&0&0&1
% \end{bmatrix}
% \begin{bmatrix}
% 1&0&0&0\\
% 0&c_{{\omega_2}}&s_{{\omega_2}}e^{-i{\phi_2}}&0\\
% 0&-s_{{\omega_2}}e^{i{\phi_2}}&c_{{\omega_2}}&0\\
% 0&0&0&1
% \end{bmatrix}
% \\
% \equiv
%  \begin{bmatrix}
% c_{{\omega_2}}&0&s_{{\omega_2}}e^{-i{\phi_2}}&0\\
% 0&1&0&0\\
% -s_{{\omega_2}}e^{i{\phi_2}}&0&c_{{\omega_2}}&0\\
% 0&0&0&1
% \end{bmatrix}
%  \begin{bmatrix}
%  0&1&0&0\\
%  -1&0&0&0\\
%  0&0&1&0\\
%  0&0&0&1
% \end{bmatrix}
%  \,.
%\end{equation}
%This is to say $\widetilde{R}_{12} R_{23}({\omega_2},{\phi_2})\approx R_{13}({\omega_2},{\phi_2})\widetilde{R}_{12}$.
%Therefore, we can use such an approximation within a different $E$ range:

In order to maintain the similar oscillation expression as much as its form in vacuum, we would like to replace $\widetilde{R}_{12} R_{23}(\omega_2,\phi_2)$ with 
\begin{equation}
\label{eq:NeuExch2}
\widetilde{R}_{12} R_{23}({\omega_2},{\phi_2})\approx R_{13}({\omega_2},{\phi_2})\widetilde{R}_{12} \,,
\end{equation}
where $\widetilde{R}_{12}R_{23}(\omega_2,\phi_2)$ is expressed as
\begin{equation}
\widetilde{R}_{12}R_{23}(\omega_2,\phi_2)=
\begin{bmatrix}
\widetilde{c}_{12} e^{i\Theta_{12}} & c_{\omega_2} \widetilde{s}_{12}e^{-i(\widetilde{\delta}_{12}+\Theta_{12})} & \widetilde{s}_{12}s_{\omega_2}e^{-i(\widetilde{\delta}_{12}+\Theta_{12}+\phi_2)} & 0 \\
-\widetilde{s}_{12}e^{i(\widetilde{\delta}_{12}+\Theta_{12})} & \widetilde{c}_{12}c_{\omega_2}e^{-i\Theta_{12}} & \widetilde{c}_{12}s_{\omega_2} e^{-i(\Theta_{12}+\phi_2)}
& 0 \\
0 & -s_{\omega_2}e^{i\phi_2} & c_{\omega_2} & 0 \\
0 & 0 & 0 & 1
\end{bmatrix}
\,,
\end{equation}
and $R_{13}(\omega_2,\phi_2)\widetilde{R}_{12}$ yields
\begin{equation}
R_{13}(\omega_2,\phi_2)\widetilde{R}_{12}=
\begin{bmatrix}
\widetilde{c}_{12}c_{\omega_2}e^{i\Theta_{12}} & c_{\omega_2}\widetilde{s}_{12} e^{-i(\widetilde{\delta}_{12}+\Theta_{12})} & s_{\omega_2}e^{-i\phi_2} & 0 \\
-\widetilde{s}_{12}e^{i(\widetilde{\delta}_{12}+\Theta_{12})} & \widetilde{c}_{12}e^{-i\Theta_{12}} & 0 & 0 \\
-\widetilde{c}_{12}s_{\omega_2} e^{i(\Theta_{12}+\phi_2)} & -\widetilde{s}_{12}s_{\omega_2}e^{i(\phi_2-\Theta_{12}-\widetilde{\delta}_{12})} & c_{\omega_2} & 0 \\
0 & 0 & 0 & 1
\end{bmatrix}\,.
\end{equation}

\begin{figure}[htbp]
\centering
\includegraphics[width=0.6\textwidth]{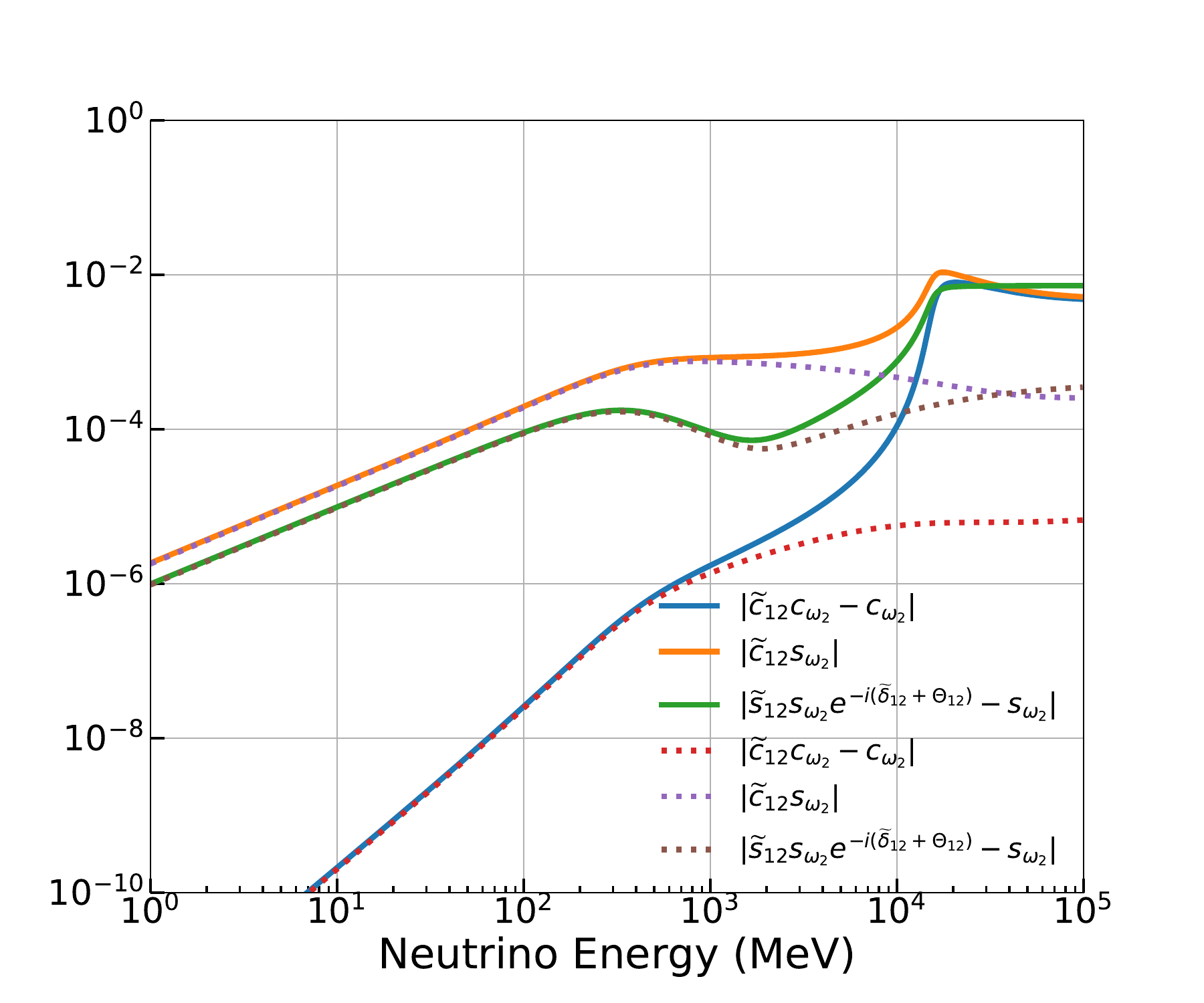}
\caption{\label{fig:6}  The differences between $\widetilde{R}_{12}R_{23}(\omega_2,\phi_2)$ and $R_{13}(\omega_2,\phi_2)\widetilde{R}_{12}$. The solid and
dashed lines are for NH and IH. The input oscillation parameters are listed in Table~\ref{table:i}.}
\end{figure}

In order to check the validity of eq.~\ref{eq:NeuExch2}, we compare all the corresponding elements between $\widetilde{R}_{12}R_{23}(\omega_2,\phi_2)$ and $R_{13}(\omega_2,\phi_2)\widetilde{R}_{12}$. As shown in figure~\ref{fig:6}, the difference between those two matrices are quite small in our application range.  
 
\begin{figure}[htbp]
\centering
\includegraphics[width=0.6\textwidth]{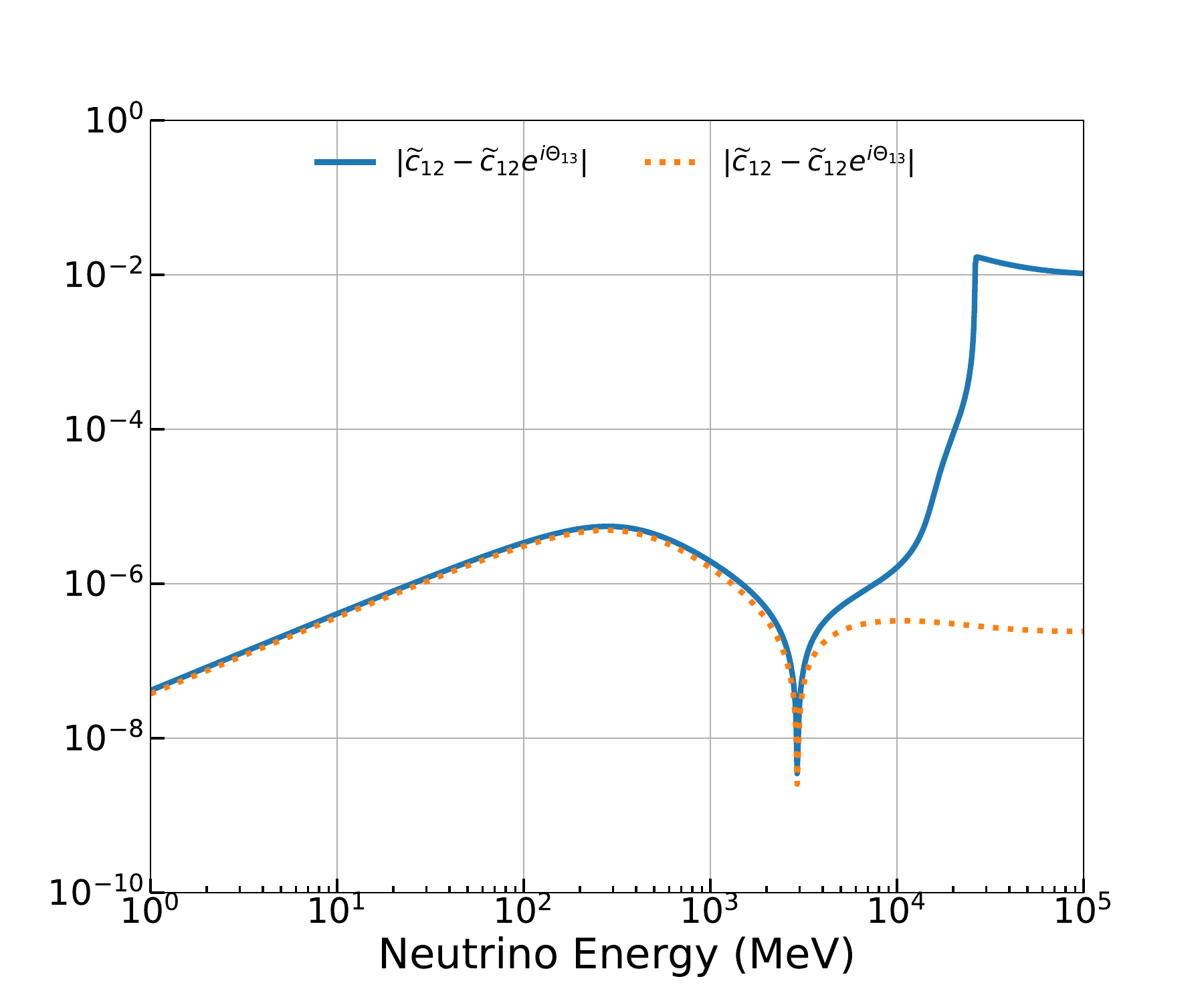}
\caption{\label{fig:7}  The differences between $D_{13}(e^{i\Theta_{13}},1,e^{-i\Theta_{13}},1)\widetilde{R}_{12}$ and $\widetilde{R}_{12} D_{23}(1,e^{i\Theta_{13}},e^{-i\Theta_{13}},1)$. The solid and dashed lines are for NH and IH. The input oscillation parameters are listed in Table~\ref{table:i}.}
\end{figure}

Subsequently, we obtain
\begin{equation}
\label{eq:NeuU1}
\widetilde{U}\approx  R_{34}R_{24}R_{14}R_{23}R_{13}\widetilde{R}_{12}R_{23}({\omega_2},{\phi_2})
=R_{34}R_{24}R_{14}R_{23}R_{13}R_{13}({\omega_2},{\phi_2})\widetilde{R}_{12} \,,
\end{equation}
where
\begin{equation}
\label{eq:NeuExch3}
\begin{split}
&R_{13}(\theta_{13},0)R_{13}({\omega_2},{\phi_2})\\
=&
\begin{bmatrix}
c_{13}&0&s_{13}e^{-i\delta_{13}}&0\\
0&1&0&0\\
-s_{13}e^{i\delta_{13}}&0&c_{13}&0\\
0&0&0&1
\end{bmatrix}
\begin{bmatrix}
c_{{\omega_2}}&0&s_{{\omega_2}}e^{-i{\phi_2}}&0\\
0&1&0&0\\
-s_{{\omega_2}}e^{i{\phi_2}}&0&c_{{\omega_2}}&0\\
0&0&0&1
\end{bmatrix}
\\=&
\begin{bmatrix}
\widetilde{c}_{13}&0&\widetilde{s}_{13}e^{-i\widetilde{\delta}_{13}}&0\\
0&1&0&0\\
-\widetilde{s}_{13}e^{i\widetilde{\delta}_{13}}&0&\widetilde{c}_{13}&0\\
0&0&0&1
\end{bmatrix}
\begin{bmatrix}
e^{i\Theta_{13}}&0&0&0\\
0&1&0&0\\
0&0&e^{-i\Theta_{13}}&0\\
0&0&0&1
\end{bmatrix}
\\=&
R_{13}(\widetilde{\theta}_{13},\widetilde{\delta}_{13})
D_{13}(e^{i\Theta_{13}},1,e^{-i\Theta_{13}},1)
\end{split} \,.
\end{equation}
The items above can be written as
\begin{subequations}\label{eq:NeuNewPar13}
\begin{gather}
\label{eq:NeuNewPar13:1}
\widetilde{s}_{13}=
\sin \widetilde{\theta}_{13}  =
\frac{|c_{13}\tan {\omega_2} e^{i{\phi_2}}+s_{13}e^{i\delta_{13}}|}
{\sqrt{1+\tan^2 {\omega_2} }} \,,
\quad
\widetilde{c}_{13}=
\cos \widetilde{\theta}_{13}  =
\frac{|c_{13}-s_{13}\tan {\omega_2} e^{i(\delta_{13}-{\phi_2})}|}
{\sqrt{1+\tan^2 {\omega_2} }} \,,
\\
\label{eq:NeuNewPar13:2}
e^{i\widetilde{\delta}_{13}}
=\frac{(c_{13}\tan {\omega_2} e^{i{\phi_2}}
+s_{13}e^{i\delta_{13}})(c_{13}-s_{13}\tan {\omega_2} e^{i(\delta_{13}-{\phi_2})})}
{\cos \widetilde{\theta}_{13} \sin \widetilde{\theta}_{13} (1+\tan^2 {\omega_2} )} \,,
\\
\label{eq:NeuNewPar13:3}
e^{i\Theta_{13}}=\frac{c_{13}-s_{13}\tan{\omega_2} e^{-i(\delta_{13}-{\phi_2})}}
{\cos\widetilde{\theta}_{13}\sqrt{1+\tan^2{\omega_2}}} \,.
\end{gather}
\end{subequations}
Equally, we set limit on $\widetilde{\theta}_{13}$ with $[0,\frac{\pi}{2}]$.
We find
\begin{equation}\label{eq:NeuU2}
\widetilde{U}\approx
R_{34}R_{24}R_{14}R_{23}
R_{13}(\widetilde{\theta}_{13},\widetilde{\delta}_{13})D_{13}\widetilde{R}_{12} \,.
\end{equation}
For the same simplification purpose, we want to replace $D_{13} \widetilde{R}_{12}$ by
\begin{equation}
\label{eq:NeuExch4}
D_{13}(e^{i\Theta_{13}},1,e^{-i\Theta_{13}},1)
\widetilde{R}_{12}
\approx
\widetilde{R}_{12}
D_{23}(1,e^{i\Theta_{13}},e^{-i\Theta_{13}},1) \,,
\end{equation}
where
\begin{equation}
D_{13}(e^{i\Theta_{13}},1,e^{-i\Theta_{13}},1)
\widetilde{R}_{12}=
\begin{bmatrix}
\widetilde{c}_{12}e^{i(\Theta_{12}+\Theta_{13})} & \widetilde{s}_{12}e^{i(\Theta_{13}-\widetilde{\delta}_{12}-\Theta_{12})} & 0 & 0 \\
-\widetilde{s}_{12} e^{i(\widetilde{\delta}_{12}+\Theta_{12})} & \widetilde{c}_{12}e^{-i\Theta_{12}} & 0 & 0 \\
0 & 0 & e^{-i\Theta_{13}} & 0 \\
0 & 0 & 0 & 1
\end{bmatrix}
\end{equation}
and 
\begin{equation}
\widetilde{R}_{12} D_{23}(1,e^{i\Theta_{13}},e^{-i\Theta_{13}},1)=
\begin{bmatrix}
\widetilde{c}_{12} e^{i\Theta_{12}} & \widetilde{s}_{12} e^{i(\Theta_{13}-\widetilde{\delta}_{12}-\Theta_{12})} & 0 & 0 \\
-\widetilde{s}_{12}e^{i(\widetilde{\delta}_{12}+\Theta_{12})} & \widetilde{c}_{12} e^{i(\Theta_{13}-\Theta_{12})} & 0 & 0 \\
0 & 0 & e^{-\Theta_{13}} & 0 \\
0 & 0 & 0 & 1
\end{bmatrix}\,.
\end{equation}
Figure~\ref{fig:7} quantifies the difference between $D_{13}(e^{i\Theta_{13}},1,e^{-i\Theta_{13}},1)\widetilde{R}_{12}$ and $\widetilde{R}_{12} D_{23}(1,e^{i\Theta_{13}},e^{-i\Theta_{13}},1)$. It is allowable to have such operation due to the small difference.

Then we obtain
\begin{equation}
\label{eq:NeuU3}
\widetilde{U}\approx
R_{34}R_{24}R_{14}R_{23}
R_{13}(\widetilde{\theta}_{13},\widetilde{\delta}_{13})
R_{12}(\widetilde{\theta}_{12},\widetilde{\delta}_{12}) D_{123}
 \,,
\end{equation}
with
\begin{equation}
\label{eq:NeuU4}
D_{123}=D_{123}(e^{i\Theta_{12}},e^{i(\Theta_{13}-\Theta_{12})},e^{-i\Theta_{13}},1) \,.
\end{equation}
Here $D_{123}(e^{i\Theta_{12}},e^{i(\Theta_{13}-\Theta_{12})},e^{-i\Theta_{13}},1)$ 
can be cancelled in the neutrino oscillation paradigm like Majorana phases. Finally we obtain the last expression for the neutrino mixing matrix:
\begin{equation}
\label{eq:NeuU5}
\widetilde{U}\approx
R_{34}R_{24}R_{14}R_{23}
R_{13}(\widetilde{\theta}_{13},\widetilde{\delta}_{13})
R_{12}(\widetilde{\theta}_{12},\widetilde{\delta}_{12}) \,,
\end{equation}
which has almost the same form with the standard mixing matrix $U$
except for an additional $\widetilde{\delta}_{12}$.
We conclude that if $\delta_{13}=\delta_{24} =\delta_{34}=0$, we can get $\widetilde{\theta}_{12}= \theta_{12}+\omega_1$,
$\widetilde{\theta}_{13}= \theta_{13}+{\omega_2}$, $\widetilde{\delta}_{13}= \delta_{13}$ and $\widetilde{\delta}_{12}=\Theta_{12}=\Theta_{13}=0$.
The effective Hamiltonian, meanwhile, becomes a real Hermitian matrix.
In that case, our method reduces to the Jacobi method, which is a way to deal with the diagonalization of real Hermitian matrices.
 \subsection{Antineutrino case}
 \label{sec:AntiNeutrinoCase}
 In this subsection, we diagonalize the effective Hamiltonian in matter and simplify the expressions of the effective mixing and mass-squared differences for the antineutrino case.
\subsubsection{Diagonalization process}
\label{sec:AntiNeutrinoCaseRotations}
In the case of antineutrinos, we rotate the 1-2 submatrix first, similar to the neutrino case.

\paragraph{First rotation:} The rotation is 
 \begin{equation}
 \label{eq:AntiNeuRoStep1}
R^1=R_{12}(\theta,{\phi_1})\equiv \begin{bmatrix}
c_{{\omega_1}}&s_{{\omega_1}}e^{-i{\phi_1}}&0&0\\
-s_{{\omega_1}}e^{i{\phi_1}}&c_{{\omega_1}}&0&0\\
0&0&1&0\\
0&0&0&1
\end{bmatrix}\,, \qquad (c_{{\omega_1}}=\cos {\omega_1} ,s_{{\omega_1}}=\sin {\omega_1})
\end{equation}
where ${\omega_1}$ is compatible with
\begin{equation}
\label{eq:AntiNeuTan1}
\tan {\omega_1} =
\frac{2A_{{\omega_1}}}{(H_{22}-H_{11})+
\sqrt{(H_{22}-H_{11})^2+4A_{{\omega_1}}^2}} \,,\qquad (0>{\omega_1}>-\theta_{12})
\end{equation}
with $A_{{\omega_1}}=-|H_{12}|$ and the additional complex factor $\phi_1=\mathrm{Arg}(\mathrm{sign}(A_{\omega_1})H_{12}^{*})$.
After the first rotation we get the new effective Hamiltonian
\begin{equation}
\label{eq:AntiNeuHPrime}
H^\prime = R_{12}^\dag({\omega_1},{\phi_1}) H R_{12}({\omega_1},{\phi_1})
=
\frac{1}{2E}\begin{bmatrix}
\lambda_{-} &0 &H_{13}^{\prime} & H_{14}^{\prime } \\
0 & \lambda_{+} & H_{23}^{\prime } & H_{24}^{\prime} \\
H_{31}^{\prime } & H_{32}^{\prime } & H_{33} & H_{34} \\
H_{41}^{\prime } & H_{42}^{\prime } & H_{43}& H_{44}
\end{bmatrix} \,.
\end{equation}
The eigenvalues of $H^\prime$ in 1-2 submatrix are
\begin{equation}\label{eq:AntiNeuHPrimeEle}
\lambda_-=\frac{H_{11}+H_{22}\tan^2 {\omega_1} -2A_{\omega_1}\tan {\omega_1} }
{1+\tan^2 {\omega_1} } \,,
\quad
\lambda_+=\frac{H_{11}\tan^2 {\omega_1} +H_{22}+2A_{\omega_1}\tan {\omega_1} }
{1+\tan^2 {\omega_1} } \,.
\end{equation}
The off-diagonal terms in $H^\prime$ are
\begin{equation}
H_{13}^\prime = H_{31}^{\prime *}=c_{\omega_1}H_{13}-s_{\omega_1} H_{23}e^{-i\phi_1}\,,
\end{equation}
\begin{equation}
H_{23}^\prime=H_{32}^{\prime *} = c_{\omega_1}H_{23} + s_{\omega_1} H_{13}e^{i\phi_1}\,,
\end{equation}
\begin{equation}
H_{14}^\prime = H_{41}^{\prime *} = c_{\omega_1}H_{14} - s_{\omega_1} H_{24}e^{-i\phi_1}\,,
\end{equation}
\begin{equation}
H_{24}^\prime= H_{42}^{\prime *}= c_{\omega_1}H_{24} + s_{\omega_1} H_{14}e^{i\phi_1}\,.
\end{equation}

\paragraph{Second rotation:} 
The rotation matrix is
\begin{equation}
\label{eq:AntiNeuRoStep2}
R^2=R_{13}({\omega_2},{\phi_2}) \equiv \begin{bmatrix}
c_{{\omega_2}}&0&s_{{\omega_2}}e^{-i{\phi_2}}&0\\
0&1&0&0\\
-s_{{\omega_2}}e^{i{\phi_2}}&0&c_{{\omega_2}}&0\\
0&0&0&1
\end{bmatrix}\,.\qquad (c_{{\omega_2}}=\cos {\omega_2} \,, s_{{\omega_2}}=\sin {\omega_2})
\end{equation}
The rotation angle ${\omega_2}$ yields
\begin{equation}
\label{eq:AntiNeuTan2}
\tan {\omega_2}
=
\frac{2A_{\omega_2}}{(H_{33}-\lambda_-)\pm
\sqrt{(H_{33}-\lambda_-)^2+4A_{\omega_2}^2}} \,,
\end{equation}
where the upper sign is for NH $( 0 > {\omega_2} > -\theta_{13}) $,
and the lower sign is for IH $ (0 < {\omega_2} < \frac{\pi}{2}-\theta_{13}) $ with $A_{{\omega_2}}=-|H_{13}^\prime|$ and a corresponding complex factor $\phi_2=\mathrm{Arg}(\mathrm{sign}(A_{\omega_2})H_{13}^{\prime*})$.
After two operations, we get the new effective Hamiltonian in the new representation:
 \begin{equation}
 \label{eq:AntiNeuHPrimePrime}
H^{\prime\prime}=  R_{13}^\dag({\omega_2},{\phi_2}) H^{\prime} R_{13}({\omega_2},{\phi_2})
=
\frac{1}{2E}\begin{bmatrix}
\lambda_{-}^\prime&H_{12}^{\prime\prime}&0&H_{14}^{\prime\prime}\\
H_{21}^{\prime\prime}&\lambda_+&H_{23}^{\prime\prime}&H_{24}^\prime\\
0&H_{32}^{\prime\prime}&\lambda_{+}^{\prime}&H_{34}^{\prime\prime}\\
H_{41}^{\prime\prime}&H_{42}^\prime&H_{43}^{\prime\prime}&H_{44}
\end{bmatrix} \,,
\end{equation}
with
\begin{equation}
\lambda_-^\prime=\frac{\lambda_-
+H_{33}\tan^2 {\omega_2} -2A_{\omega_2}\tan {\omega_2} }
{1+\tan^2 {\omega_2} } \,,
\quad
\lambda_+^\prime=\frac{\lambda_-\tan^2 {\omega_2} +H_{33}
+2A_{\omega_2}\tan {\omega_2} }
{1+\tan^2 {\omega_2} } \,.
\end{equation}
The residual off-diagonal terms in $H^{\prime\prime}$ are 
\begin{equation}
H^{\prime\prime}_{12}=H^{\prime\prime*}_{21}= - s_{\omega 2} H_{32}^\prime e^{-i\phi_2} \,,
\end{equation}
\begin{equation}
H^{\prime\prime}_{23}=H^{\prime\prime*}_{32}=c_{\omega 2} H_{23}^\prime  \,,
\end{equation}
\begin{equation}
H^{\prime\prime}_{14}=H^{\prime\prime*}_{41}= c_{\omega 2} H_{14}^\prime - s_{\omega 2} H_{34} e^{-i\phi_2} \,,
\end{equation}
\begin{equation}
H^{\prime\prime}_{34}=H^{\prime\prime*}_{43}= c_{\omega 2} H_{34} + s_{\omega 2} H_{14}^\prime e^{i\phi_2} \,,
\end{equation}

\begin{figure}[htbp]
\centering
\includegraphics[width=0.6\textwidth]{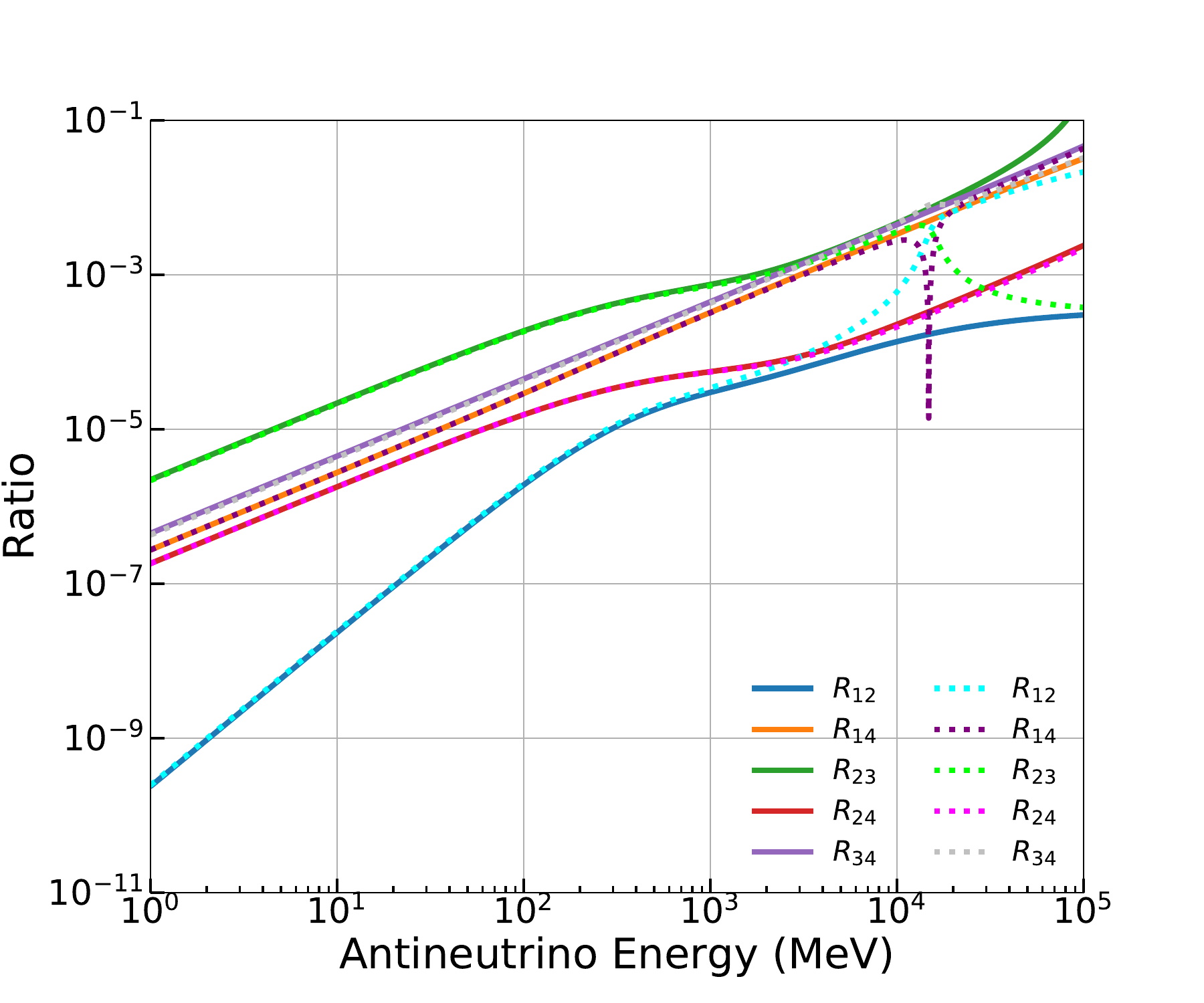}
\caption{\label{fig:8}  The values of the off-diagonal elements in $H^{\prime\prime}$. The solid and dashed lines are for NH and IH. The input oscillation parameters are listed in Table~\ref{table:i}.}
\end{figure}

% with some negligible off-diagonal elements in
%$H^{\prime\prime}$ \eqref{eq:AntiNeuHPrimePrime}.
Figure~\ref{fig:8} shows that the off-diagonal terms of $H^{\prime\prime}$ are negligible as a good approximation.
So far we get the effective mixing matrix $\widetilde{U}$ and
effective mass-squared $\Delta \widetilde{m}_{i1}^2 (i=2,3,4)$.
Using the above rotations, we obtain
 \begin{subequations}
\label{eq:EffAntiNeuDU}
\begin{gather}
\label{eq:EffAntiNeuDU:1}
\widetilde{U}\approx UR_{12}({\omega_1},{\phi_1})R_{23}({\omega_2},{\phi_2})
=\underbrace{R_{34}R_{24}R_{14}R_{23}R_{13}R_{12}}_{U}
R_{12}({\omega_1},{\phi_1})R_{13}({\omega_2},{\phi_2}) \,,
\\
\label{eq:EffAntiNeuDU:2}
\Delta \widetilde{m}_{21}^2\approx\lambda_+-\lambda_-^\prime \,,
\quad
\Delta \widetilde{m}_{31}^2\approx\lambda_{+}^{\prime}-\lambda_-^\prime \,,
\quad
\Delta \widetilde{m}_{41}^2\approx H_{44}-\lambda_-^\prime \,.
\end{gather}
\end{subequations}
\subsubsection{Simplification}
 \label{sec:AntiNeutrinoCaseAbsorpation}
Similar to the neutrino case, we combine $R_{12}(\theta_{12},0)$ and $R_{12}({\omega_1},{\phi_1})$
into one submatrix below:
 \begin{equation}
 \label{eq:EffAntiNeuR12}
\begin{split}
&R_{12}(\theta_{12},0)R_{12}({\omega_1},{\phi_1})\\
=&
\begin{bmatrix}
c_{12}&s_{12}&0&0\\
-s_{12}&c_{12}&0&0\\
0&0&1&0\\
0&0&0&1
\end{bmatrix}
\begin{bmatrix}
c_{{\omega_1}}&s_{{\omega_1}}e^{-i{\phi_1}}&0&0\\
-s_{{\omega_1}}e^{i{\phi_1}}&c_{{\omega_1}}&0&0\\
0&0&1&0\\
0&0&0&1
\end{bmatrix}
\\=&
\begin{bmatrix}
\widetilde{c}_{12}&\widetilde{s}_{12}e^{-i\widetilde{\delta}_{12}}&0&0\\
-\widetilde{s}_{12}e^{i\widetilde{\delta}_{12}}&\widetilde{c}_{12}&0&0\\
0&0&1&0\\
0&0&0&1
\end{bmatrix}
\begin{bmatrix}
e^{i\Theta_{12}}&0&0&0\\
0&e^{-i\Theta_{12}}&0&0\\
0&0&1&0\\
0&0&0&1
\end{bmatrix}
\\=&
R_{12}(\widetilde{\theta}_{12},\widetilde{\delta}_{12})D_{12}(e^{i\Theta_{12}},e^{-i\Theta_{12}},1,1)
\\=&
\widetilde{R}_{12}
\end{split} \,,
\end{equation}
with
\begin{subequations}\label{eq:AntiNeuNewPar12}
\begin{gather}
\label{eq:AntiNeuNewPar12:1}
\widetilde{s}_{12}=
\sin \widetilde{\theta}_{12}
 = \frac{|c_{12}\tan {\omega_1} e^{i{\phi_1}}+s_{12}|}{\sqrt{1+\tan^2 {\omega_1} }}\,,
\quad
\widetilde{c}_{12}=
\cos \widetilde{\theta}_{12}
 = \frac{|c_{12}-s_{12}\tan {\omega_1} e^{i{\phi_1}}|}{\sqrt{1+\tan^2 {\omega_1} }}\,,
\\
\label{eq:AntiNeuNewPar12:2}
e^{i\widetilde{\delta}_{12}}
=
\frac{(c_{12}\tan {\omega_1} e^{i{\phi_1}}+s_{12})(c_{12}-s_{12}\tan {\omega_1} e^{-i{\phi_1}})}
{\cos \widetilde{\theta}_{12} \sin \widetilde{\theta}_{12} (1+\tan^2 {\omega_1}) } \,,
\\
\label{eq:AntiNeuNewPar12:3}
e^{i\Theta_{12}}=\frac{c_{12}-s_{12}\tan{\omega_1} e^{\phi_1}}{\cos\widetilde{\theta}_{12}\sqrt{1+\tan^2{\omega_1}}}
\,.
\end{gather}
\end{subequations}
Therefore, we get 
\begin{equation}
\widetilde{U}\approx  R_{34}R_{24}R_{14}R_{23}R_{13}\widetilde{R}_{12}R_{13}({\omega_2},{\phi_2}) \,.
\end{equation}
Here, we set a constraint on $\widetilde{\theta}_{12}$ within $[0,\frac{\pi}{2}]$.
% After the absorption of two rotations in the first step, we get an effective $\widetilde{R}_{12}$. 
% The energy dependent $\widetilde{\theta}_{12}$ is shown in figure~\ref{fig:1}.
% When $E$ is small, matter effects are slight, leading to $\omega_2\rightarrow 0$. 
% Therefore, we get $\widetilde{R}_{12}  R_{13}({\omega_2},{\phi_2})\approx R_{13}({\omega_2},{\phi_2})\widetilde{R}_{12}$. 
% When $E$ goes up, $\cos\theta_{12}\rightarrow 1$, leading to a unitary matrix $\widetilde{R}_{12}$
%\begin{equation} \label{eq:AntiNeuExch}
%\widetilde{R}_{12} \approx
% \begin{bmatrix}
%  1&0&0&0\\
%  0&1&0&0\\
%  0&0&1&0\\
%  0&0&0&1
% \end{bmatrix} \,.
%\end{equation}

For further simplification, we exchange $\widetilde{R}_{12}$ and  $R_{13}({\omega_2},{\phi_2})$ by
\begin{equation}
\widetilde{R}_{12}  R_{13}({\omega_2},{\phi_2})\approx R_{13}({\omega_2},{\phi_2})\widetilde{R}_{12}
\end{equation}
where
\begin{equation}
\widetilde{R}_{12} R_{13}({\omega_2},{\phi_2}) =
\begin{bmatrix}
\widetilde{c}_{12}c_{\omega_2}e^{i\Theta_{12}} & \widetilde{s}_{12}e^{-i(\widetilde{\delta}_{12}+\Theta_{12})} & \widetilde{c}_{12}s_{\omega_2} e^{i (\Theta_{12}-\phi_2 )} & 0 \\
-c_{\omega_2}\widetilde{s}_{12}e^{i(\widetilde{\delta}_{12}+\Theta_{12})} & \widetilde{c}_{12}e^{-i\Theta_{12}} & -\widetilde{s}_{12}s_{\omega_2}e^{i(\widetilde{\delta}_{12} + \Theta_{12} - \phi_2)} & 0 \\
-s_{\omega_2}e^{i\phi_2} & 0 & c_{\omega_2} & 0 \\
0 & 0 & 0 & 1
\end{bmatrix}\,,
\end{equation}
and $R_{13}(\omega_2,\phi_2)\widetilde{R}_{12}$ yields
\begin{equation}
R_{13}(\omega_2,\phi_2)\widetilde{R}_{12}=
\begin{bmatrix}
\widetilde{c}_{12}c_{\omega_2}e^{i\Theta_{12}} & c_{\omega_2}\widetilde{s}_{12} e^{-i(\widetilde{\delta}_{12}+\Theta_{12})} & s_{\omega_2}e^{-i\phi_2} & 0 \\
-\widetilde{s}_{12}e^{i(\widetilde{\delta}_{12}+\Theta_{12})} & \widetilde{c}_{12}e^{-i\Theta_{12}} & 0 & 0 \\
-\widetilde{c}_{12}s_{\omega_2} e^{i(\Theta_{12}+\phi_2)} & -\widetilde{s}_{12}s_{\omega_2}e^{i(\phi_2-\Theta_{12}-\widetilde{\delta}_{12})} & c_{\omega_2} & 0 \\
0 & 0 & 0 & 1
\end{bmatrix}\,.
\end{equation}

\begin{figure}[htbp]
\centering
\includegraphics[width=0.6\textwidth]{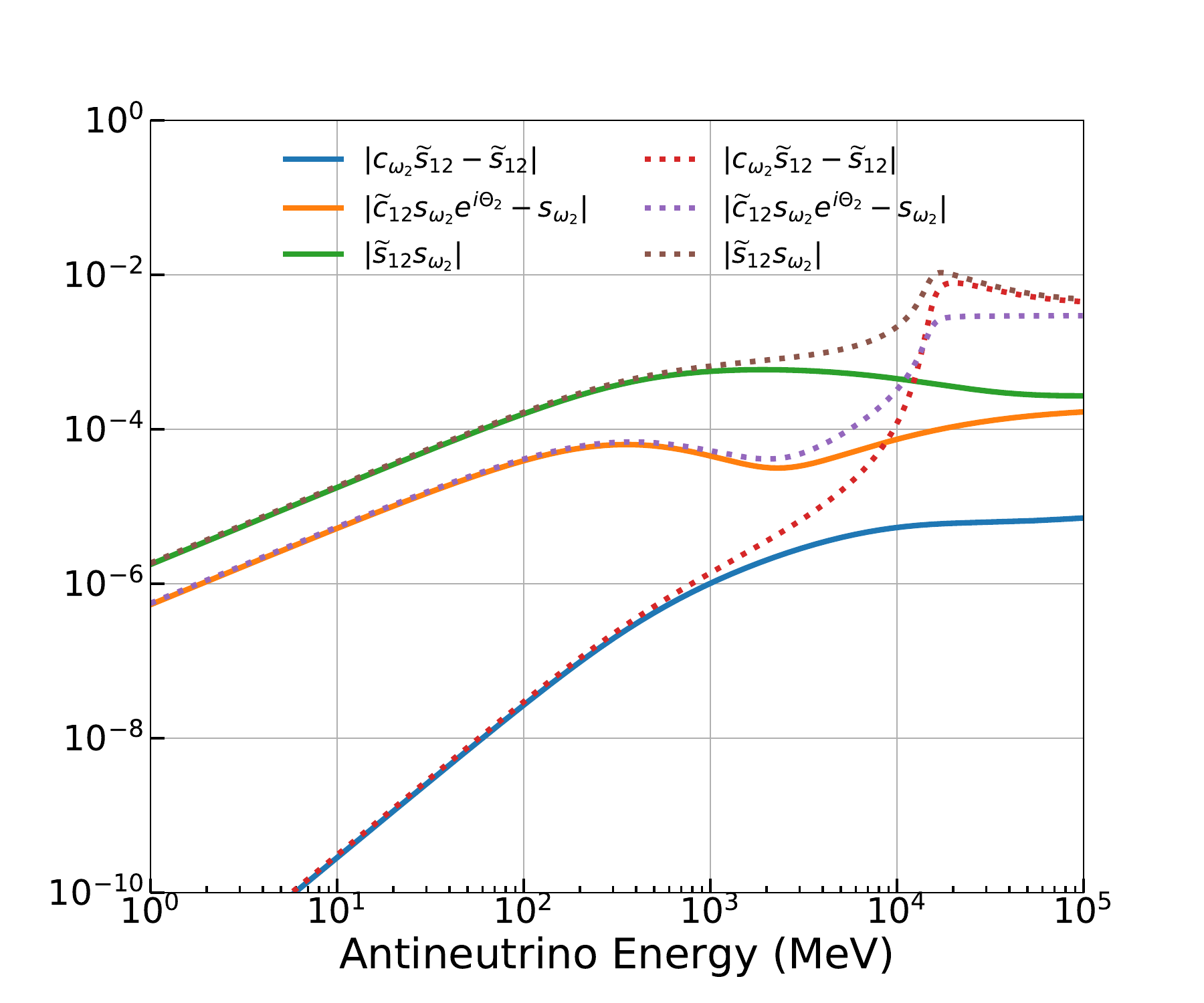}
\caption{\label{fig:9}  The differences between $\widetilde{R}_{12}R_{23}(\omega_2,\phi_2)$ and $R_{13}(\omega_2,\phi_2)\widetilde{R}_{12}$. The solid and
dashed lines are for NH and IH. The input oscillation parameters are listed in Table~\ref{table:i}.}
\end{figure}

Figure~\ref{fig:9} quantifies the differences of all elements between $\widetilde{R}_{12} R_{13}(\omega_2,\phi_2)$ and $R_{13}(\omega_2,\phi_2)\widetilde{R}_{12}$. It shows that the differences are allowable for exchange in the application range of our approximation.

%Then we get $\widetilde{R}_{12}  R_{13}({\omega_2},{\phi_2})\approx R_{13}({\omega_2},{\phi_2})\widetilde{R}_{12}$ as well. 
Subsequently, we obtain the following approximate simplification over different $E$ ranges: 
\begin{equation}\label{eq:AntiNeuU1}
\widetilde{U}\approx  R_{34}R_{24}R_{14}R_{13}R_{13}\widetilde{R}_{12}  R_{13}({\omega_2},{\phi_2})
=R_{34}R_{24}R_{14}R_{13}R_{13}R_{13}({\omega_2},{\phi_2})\widetilde{R}_{12}.
\end{equation}
 We notice that $R_{13}(\theta_{13},0)R_{13}({\omega_2},{\phi_2})$ can be simplified as
\begin{equation}\label{eq:AntiNeuExch2}
\begin{split}
&R_{13}(\theta_{13},0)R_{13}({\omega_2},{\phi_2})\\
=&
\begin{bmatrix}
c_{13}&0&s_{13}e^{-i\delta_{13}}&0\\
0&1&0&0\\
-s_{13}e^{i\delta_{13}}&0&c_{13}&0\\
0&0&0&1
\end{bmatrix}
\begin{bmatrix}
c_{{\omega_2}}&0&s_{{\omega_2}}e^{-i{\phi_2}}&0\\
0&1&0&0\\
-s_{{\omega_2}}e^{i{\phi_2}}&0&c_{{\omega_2}}&0\\
0&0&0&1
\end{bmatrix}
\\=&
\begin{bmatrix}
\widetilde{c}_{13}&0&\widetilde{s}_{13}e^{-i\widetilde{\delta}_{13}}&0\\
0&1&0&0\\
-\widetilde{s}_{13}e^{i\widetilde{\delta}_{13}}&0&\widetilde{c}_{13}&0\\
0&0&0&1
\end{bmatrix}
\begin{bmatrix}
e^{i\Theta_{13}}&0&0&0\\
0&1&0&0\\
0&0&e^{-i\Theta_{13}}&0\\
0&0&0&1
\end{bmatrix}
\\=&
R_{13}(\widetilde{\theta}_{13},\widetilde{\delta}_{13})D_{13}(e^{i\Theta_{13}},1,e^{-i\Theta_{13}},1)
\end{split} \,,
\end{equation}
 with
\begin{subequations}\label{eq:AntiNeuNewPar13}
\begin{gather}
\label{eq:AntiNeuNewPar13:1}
\widetilde{s}_{13}=
\sin \widetilde{\theta}_{13}  =
\frac{|c_{13}\tan {\omega_2} e^{i{\phi_2}}+s_{13}e^{i\delta_{13}}|}
{\sqrt{1+\tan^2 {\omega_2} }}
\,,
\quad
\widetilde{c}_{13}=
\cos \widetilde{\theta}_{13}  =
\frac{|c_{13}-s_{13}\tan {\omega_2} e^{i(\delta_{13}-{\phi_2})}|}
{\sqrt{1+\tan^2 {\omega_2} }}
\,,
\\
\label{eq:AntiNeuNewPar13:2}
e^{i\widetilde{\delta}_{13}}
=
\frac{(c_{13}\tan {\omega_2} e^{i{\phi_2}}+s_{13}e^{i\delta_{13}})(c_{13}-s_{13}\tan {\omega_2} e^{i(\delta_{13}-{\phi_2})})}
{\cos \widetilde{\theta}_{13} \sin \widetilde{\theta}_{13} (1+\tan^2 {\omega_2} )}
\,,
\\
\label{eq:AntiNeuNewPar13:3}
e^{i\Theta_{13}}=\frac{c_{13}-s_{13}\tan{\omega_2} e^{-i(\delta_{13}-{\phi_2})}}{\cos\widetilde{\theta}_{13}\sqrt{1+\tan^2{\omega_2}}}
\,.
\end{gather}
\end{subequations}
We also set a bound on $\widetilde{\theta}_{13}$ within $[0,\frac{\pi}{2}]$.
Subsequently, we obtain
\begin{equation}\label{eq:AntiNeuU2}
\widetilde{U}\approx
R_{34}R_{24}R_{14}R_{13}
R_{13}(\widetilde{\theta}_{13},\widetilde{\delta}_{13})D_{13}
\widetilde{R}_{12} \,.
\end{equation}

Similar to the approach from eq.~\eqref{eq:AntiNeuU1}, we want to exchange $D_{13}(e^{i\Theta_{13}},1,e^{-i\Theta_{13}},1)$ and $\widetilde{R}_{12}$ by
\begin{equation}
D_{13}(e^{i\Theta_{13}},1,e^{-i\Theta_{13}},1)
\widetilde{R}_{12}
\approx
\widetilde{R}_{12}
D_{13}(e^{i\Theta_{13}},1,e^{-i\Theta_{13}},1) \,,
\end{equation}
where 
\begin{equation}
D_{13}(e^{i\Theta_{13}},1,e^{-i\Theta_{13}},1) \widetilde{R}_{12}=
\begin{bmatrix}
\widetilde{c}_{12}e^{i(\Theta_{12}+\Theta_{13})} & \widetilde{s}_{12}e^{i(\Theta_{13}-\widetilde{\delta}_{12}-\Theta_{12})} & 0 & 0 \\
-\widetilde{s}_{12} e^{i(\widetilde{\delta}_{12}+\Theta_{12})} & \widetilde{c}_{12}e^{-i\Theta_{12}} & 0 & 0 \\
0 & 0 & e^{-i\Theta_{13}} & 0 \\
0 & 0 & 0 & 1
\end{bmatrix}
\end{equation}
and 
\begin{equation}
\widetilde{R}_{12} D_{13}(e^{i\Theta_{13}},1,e^{-i\Theta_{13}},1) =
\begin{bmatrix}
\widetilde{c}_{12}e^{i(\Theta_{12}+\Theta_{13})} & \widetilde{s}_{12}e^{-i(\widetilde{\delta}_{12}+\Theta_{12})} & 0 & 0 \\
-\widetilde{s}_{12} e^{i(\widetilde{\delta}_{12}+\Theta_{12}+\Theta_{13})} & \widetilde{c}_{12}e^{-i\Theta_{12}} & 0 & 0 \\
0 & 0 & e^{-i\Theta_{13}} & 0 \\
0 & 0 & 0 & 1
\end{bmatrix}\,.
\end{equation}

\begin{figure}[htbp]
\centering
\includegraphics[width=0.6\textwidth]{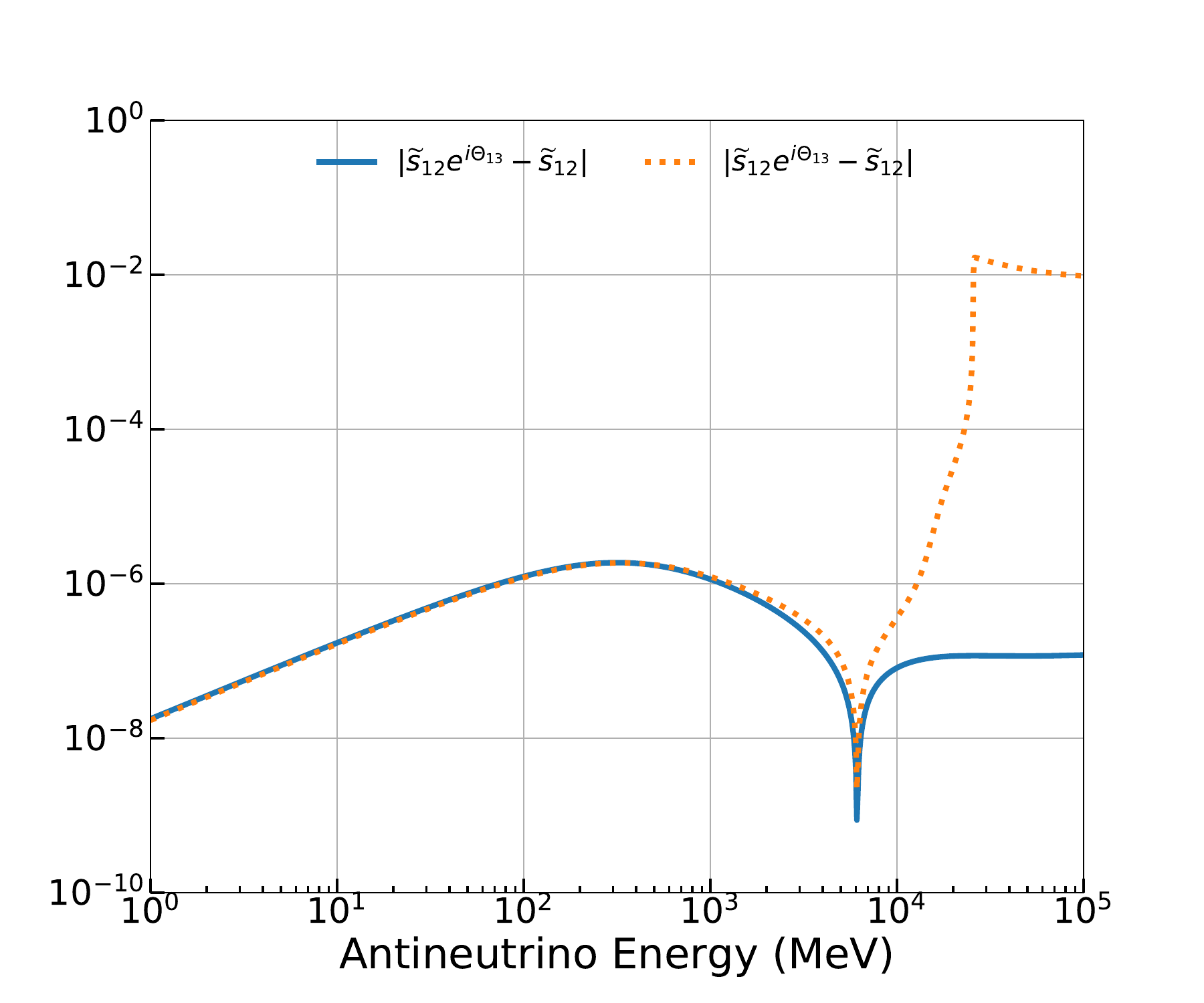}
\caption{\label{fig:10}  The differences between $D_{13}(e^{i\Theta_{13}},1,e^{-i\Theta_{13}},1)\widetilde{R}_{12}$ and $\widetilde{R}_{12} D_{13}(e^{i\Theta_{13}},1,e^{-i\Theta_{13}},1)$. The solid and dashed lines are for NH and IH. The input oscillation parameters are listed in Table~\ref{table:i}.}
\end{figure}

Figure~\ref{fig:10} quantifies the negligible difference between 
$D_{13}(e^{i\Theta_{13}},1,e^{-i\Theta_{13}},1) \widetilde{R}_{12}$
and $\widetilde{R}_{12} D_{13}(e^{i\Theta_{13}},1,e^{-i\Theta_{13}},1)$.
Again, the difference are quite small.  

Consequently, the effective mixing matrix can be written as
\begin{equation}
\widetilde{U}\approx
R_{34}R_{24}R_{14}R_{13}
R_{13}(\widetilde{\theta}_{13},\widetilde{\delta}_{13})
R_{12}(\widetilde{\theta}_{12},\widetilde{\delta}_{12}) D_{123}
\end{equation}
with
\begin{equation}
\label{eq:AntiNeuExch6}
D_{123}=D_{123}(e^{i(\Theta_{12}+\Theta_{13})},e^{-i\Theta_{12}},e^{-i\Theta_{13}},1) \,.
\end{equation}
Similarly, here $D_{123}(e^{i(\Theta_{12}+\Theta_{13})},e^{-i\Theta_{12}},e^{-i\Theta_{13}},1)
$ can be cancelled like Majorana phases when neutrinos oscillate.
Eventually, the effective mixing matrix $\widetilde{U}$ is identical to the vacuum case $U$ except for an additional $\widetilde{\delta}_{12}$.
 $\widetilde{U}$ can be written as
\begin{equation}
\label{eq:AntiNeuU3}
\begin{split}
\widetilde{U}\approx&
R_{34}R_{24}R_{14}R_{13}
R_{13}(\widetilde{\theta}_{13},\widetilde{\delta}_{13})
R_{12}(\widetilde{\theta}_{12},\widetilde{\delta}_{12}) \,.
\end{split}
\end{equation}
Similarly, We conclude that $\widetilde{\theta}_{12}=\theta_{12}+\omega_1$,
$\widetilde{\theta}_{13}=\theta_{13}+{\omega_2}$, $\widetilde{\delta}_{13}=\delta_{13}$ and
$\widetilde{\delta}_{12}=\Theta_{12}=\Theta_{13}=0$
when $\delta_{13}=\delta_{24} =\delta_{34}=0$. It is much easier to diagonalize the effective Hamiltonian than before because it is a real Hermitian matrix. For real matrices the Jacobi-like method reduces to the Jacobi method. 
\subsection{Effective mixing matrix for 3+1 flavor neutrino in matter}
\label{sec:appendixC}
In this subsection, we summarize the effective mixing matrix $\widetilde{U}$
%\footnote{It is equivalent to use $\delta_{14}$ and $\delta_{24}$, or $\delta_{14}$ and $\delta_{34}$ for the additional CP phases.} 
for both the neutrino and antineutrino cases from the results of \ref{sec:NeutrinoCase} and \ref{sec:AntiNeutrinoCase} in table~\ref{table:2}. 
Using the neutrino oscillation probability functions in eq.~\eqref{eq:EffPro} and the elements in table~\ref{table:2},
every neutrino oscillation probability is available. 
%And we also give some useful notations for calculating neutrino oscillation
%probabilities with matter effect according to section \ref{sec:Appro}.
% Then we show the accuracies of 
%some long baseline experiments using the approximations based on section~\ref{sec:Appro:EffUM}.  
%In addition, we show popular probability formulas for some accelerator experiments in matter 
%by the approximations of the third-best accuracy level.
%The best approximation of the effective
%$\widetilde{U}$ with the first-best accuracy level is given in section~\ref{sec:NeutrinoCaseRotations}
%and \ref{sec:AntiNeutrinoCaseRotations}.
\begin{table}[htbp]
 %\label{table:1}
 \renewcommand{\multirowsetup}{\centering}
 \centering
\begin{tabular}{|c|c|c|}
\hline
$\alpha$& $\widetilde{U}_{\alpha i}$ &  -\\
\hline
\multirow{4}*{$e$}&$\widetilde{U}_{e1}$&
$\widetilde{c}_{12}\widetilde{c}_{13}c_{14}$
\\
\cline{2-3}
&$\widetilde{U}_{e2}$&
$\widetilde{c}_{13}c_{14}\widetilde{s}_{12}e^{-i\widetilde{\delta}_{12}}$
\\
\cline{2-3}
&$\widetilde{U}_{e3}$&
$c_{14}\widetilde{s}_{13}e^{-i\widetilde{\delta}_{13}}$
\\
\cline{2-3}
&$\widetilde{U}_{e4}$&
$s_{14}$
\\
\hline
\multirow{4}*{$\mu$}&$\widetilde{U}_{\mu1}$&
$-\widetilde{s}_{12}c_{23}c_{24}e^{i\widetilde{\delta}_{12}}-\widetilde{c}_{12}(\widetilde{s}_{13}c_{24}s_{23}e^{i\widetilde{\delta}_{13}}+\widetilde{c}_{13}s_{14}s_{24}e^{-i\delta_{24}})$
\\
\cline{2-3}
&$\widetilde{U}_{\mu2}$&
$\widetilde{c}_{12}c_{23}c_{24}-\widetilde{s}_{12}(\widetilde{s}_{13}c_{24}s_{23}e^{i\widetilde{\delta}_{13}}+\widetilde{c}_{13}s_{14}s_{24}e^{-i\delta_{24}})e^{-i\widetilde{\delta}_{12}}$
\\
\cline{2-3}
&$\widetilde{U}_{\mu3}$&
$\widetilde{c}_{13}c_{24}s_{23}-\widetilde{s}_{13}s_{14}s_{24}e^{-i\widetilde{\delta}_{13}}e^{-i\delta_{24}}$
\\
\cline{2-3}
&$\widetilde{U}_{\mu4}$&
$c_{14}s_{24}e^{-i\delta_{24}}$
\\
\hline
\multirow{4}*{$\tau$}&$\widetilde{U}_{\tau1}$&
$\begin{aligned}
\widetilde{c}_{12}[\widetilde{s}_{13}(s_{23}s_{24}s_{34}e^{i\delta_{24}}e^{-i\delta_{34}}-&c_{23}c_{34})e^{i\widetilde{\delta}_{13}}-\widetilde{c}_{13}c_{24}s_{14}s_{34}e^{-i\delta_{34}}]
\\
\widetilde{s}_{12}(c_{34}s_{23}+c_{23}&s_{24}s_{34}e^{i\delta_{24}}e^{-i\delta_{34}})e^{i\widetilde{\delta}_{12}}
\end{aligned}$
\\
\cline{2-3}
&$\widetilde{U}_{\tau2}$&
$\begin{aligned}
\widetilde{s}_{12}[\widetilde{s}_{13}
(s_{23}s_{24}s_{34}e^{i\delta_{24}}e^{-i\delta_{34}}-&c_{23}c_{34})e^{i\widetilde{\delta}_{13}}
-\widetilde{c}_{13}c_{24}s_{14}s_{34}e^{-i\delta_{34}}]e^{-i\widetilde{\delta}_{12}}
\\
-\widetilde{c}_{12}(c_{34}s_{23}+&c_{23}s_{24}s_{34}e^{i\delta_{24}}e^{-i\delta_{34}})
\end{aligned}$\\
\cline{2-3}
&$\widetilde{U}_{\tau3}$&
$\widetilde{c}_{13}(c_{23}c_{34}-s_{23}s_{24}s_{34}e^{i\delta_{24}}e^{-i\delta_{34}})-\widetilde{s}_{13}c_{24}s_{14}s_{34}e^{-i\widetilde{\delta}_{13}}e^{-i\delta_{34}}$
\\
\cline{2-3}
&$\widetilde{U}_{\tau4}$&
$c_{14}c_{24}s_{34}e^{-i\delta_{34}}$
\\
\hline
\multirow{4}*{$s$}&$\widetilde{U}_{s1}$&
$\begin{aligned}
\widetilde{c}_{12}[\widetilde{s}_{13}(c_{34}s_{23}s_{24}e^{i\delta_{24}}+c_{23}s_{34}&e^{i\delta_{34}})e^{i\widetilde{\delta}_{13}}-\widetilde{c}_{13}c_{24}c_{34}s_{14}]+
\\
\widetilde{s}_{12}(c_{23}c_{34}s_{24}e^{i\delta_{24}}-&s_{23}s_{34}e^{i\delta_{34}})e^{i\widetilde{\delta}_{12}}
\end{aligned}$
\\
\cline{2-3}
&$\widetilde{U}_{s2}$&
$\begin{aligned}
\widetilde{s}_{12}[\widetilde{s}_{13}
(c_{34}s_{23}s_{24}e^{i\delta_{24}}+c_{23}s_{34}&e^{i\delta_{34}})e^{i\widetilde{\delta}_{13}}-\widetilde{c}_{13}c_{24}c_{34}s_{14}]e^{-i\widetilde{\delta}_{12}}
\\
+\widetilde{c}_{12}(s_{23}s_{34}e^{i\delta_{34}}-&c_{23}c_{34}s_{24}e^{i\delta_{24}})
\end{aligned}$
\\
\cline{2-3}
&$\widetilde{U}_{s3}$&
$-\widetilde{c}_{13}(c_{34}s_{23}s_{24}e^{i\delta_{24}}+c_{23}s_{34}e^{i\delta_{34}})-c_{24}c_{34}s_{14}e^{-i\widetilde{\delta}_{13}}$
\\
\cline{2-3}
&$\widetilde{U}_{s4}$&
$c_{14}c_{24}c_{34}$
\\
\hline
\end{tabular}
\caption{\label{table:2} The elements of the effective mixing matrix for the 3+1-neutrino case in matter
based on the results from section~\ref{sec:NeutrinoCase} and \ref{sec:AntiNeutrinoCase} . If we set the sterile parameters to zero, this mixing will reduce to the effective mixing matrix for 3-flavor neutrinos with matter effect.}
\end{table}

\end{document}